%% file: main.tex
\documentclass[preprint,12pt]{elsarticle}
\journal{arXiv}

\usepackage{amssymb}

\usepackage{fancyhdr}
\usepackage[hidelinks,hypertexnames=false]{hyperref}
\input{math_shortcuts.tex}
\usepackage{siunitx}
\usepackage[capitalize]{cleveref}
\usepackage{subcaption}
\usepackage[nolist]{acronym}
\input{acronyms.tex}
\biboptions{sort&compress}
\usepackage{booktabs}
\usepackage{overpic}
\usepackage[final]{changes}

\usepackage[norelsize, linesnumbered, ruled, lined, boxed, commentsnumbered]{algorithm2e}
\usepackage{adjustbox}

\usepackage{tikz}
\usetikzlibrary{external}
\usetikzlibrary{arrows.meta}
\usetikzlibrary{shapes.misc}
\usepackage{import}

\input{colors.tex}

\newcommand{\importtikz}[2]{%
  \includegraphics{img_preprint_#1_#2}
}%

\newcommand{\importpgf}[2]{%
  \includegraphics{img_preprint_#1_#2}
}%

\fancypagestyle{firststyle}
{
   \fancyhf{}
   \fancyfoot[L]{\footnotesize 
   © 2024. This manuscript version from March 29, 2024, was accepted for publication (\textit{Computer Methods in Applied Mechanics and Engineering}, Volume 426, p. 116979, \url{https://doi.org/10.1016/j.cma.2024.116979}), and is made available under the CC-BY-NC-ND 4.0 license \url{https://creativecommons.org/licenses/by-nc-nd/4.0/}}

}

\begin{document}

\title{Eulerian Formulation of the Tensor-Based Morphology Equations for Strain-Based Blood Damage Modeling}

\author[CATS]{Nico Dirkes\corref{cor}}
\ead{dirkes@aices.rwth-aachen.de}
\cortext[cor]{Corresponding author}

\author[ILSB]{Fabian Key}

\author[CATS]{Marek Behr}

\date{March 29, 2024}

\affiliation[CATS]{
  organization={Chair for Computational Analysis of Technical Systems \\ 
  RWTH Aachen University},
  addressline={Schinkelstr. 2}, 
  city={52062 Aachen},
  country={Germany}}

\affiliation[ILSB]{organization={Institute of Lightweight Design and Structural Biomechanics \\ TU Wien},
            addressline={Gumpendorfer Str. 7}, 
            city={A-1060 Vienna},
            country={Austria}}

\include{sections_abstract.tex}


\begin{frontmatter}

\end{frontmatter}
\thispagestyle{firststyle}

\input{sections_introduction.tex}
\input{sections_model.tex}
\input{sections_numerical.tex}

\input{sections_results.tex}
\input{sections_conclusion.tex}
\input{sections_acknowledgements.tex}

\appendix
\input{sections_rotation.tex}
\input{sections_implementation.tex}

\bibliographystyle{elsarticle-num-nodoi} 
\bibliography{RBC.bib}

\end{document}

%% file: math_shortcuts.tex
\usepackage{amsmath}
\usepackage{amsfonts}
\usepackage{amssymb}
\usepackage{mathtools}
\usepackage{bm}

\let\originalleft\left
\let\originalright\right
\renewcommand{\left}{\mathopen{}\mathclose\bgroup\originalleft}
\renewcommand{\right}{\aftergroup\egroup\originalright}

\newcommand{\od}[2]{\frac{\mathrm{d}#1}{\mathrm{d} #2}}
\newcommand{\oddt}[1]{\od{#1}{t}}
\newcommand{\pd}[2]{\frac{\partial#1}{\partial #2}}
\newcommand{\comm}[2]{\left[ #1 , \, #2 \right]}
\newcommand{\dist}[2]{\left( #1 , \, #2 \right)}%
\newcommand{\reco}[1]{ \left( #1 - g\left( #1 \right) \unitM \right) }
\newcommand{\transp}{^{\mathbf{T}}}

\renewcommand{\vec}[1]{ \bm{ \mathbf{ #1 } } }
\newcommand{\mat}[1]{ \bm{ \mathbf{ #1 } } }
\newcommand{\matDer}[1]{\pd{#1}{t} + (\vec u \cdot \vec \nabla) #1}
\newcommand{\symVelGrad}{\frac{1}{2} \left( \big( \vec \nabla \vec u \big) + \big( \vec \nabla \vec u \big)\transp \right)}
\newcommand{\antisymVelGrad}{\frac{1}{2} \left( \big( \vec \nabla \vec u \big) - \big( \vec \nabla \vec u \big)\transp \right)}
\newcommand{\trafoCoord}[1]{\Qm\transp #1 \Qm}

\newcommand{\zeroMat}{\mat{0}}
\newcommand{\zeroVec}{\vec{0}}
\newcommand{\domain}{\mathfrak{D}}
\newcommand{\domainr}{\mathfrak{D}^\mathrm{rot}}
\newcommand{\intMeasure}{\vec{x}}

\newcommand{\morph}{\mat{S}}

\newcommand{\strain}{\mat{E}}
\newcommand{\straint}{\tilde{\mat{E}}}
\newcommand{\strainh}{\hat{\mat{E}}}
\newcommand{\strainI}[1]{E_{#1}}
\newcommand{\straintI}[1]{\tilde{E}_{#1}}

\newcommand{\vort}{\mat{W}}
\newcommand{\vortt}{\tilde{\mat{W}}}
\newcommand{\vortI}[1]{W_{#1}}
\newcommand{\vorttI}[1]{\tilde{W}_{#1}}

\newcommand{\unitM}{\mat{I}}

\newcommand{\Om}{\mat{\Omega}}
\newcommand{\Omt}{\tilde{\mat{\Omega}}}

\newcommand{\Omr}{\mat{\Omega}^\mathrm{imp}}

\newcommand{\OmtI}[1]{\tilde{\Omega}_{#1}}

\newcommand{\Qm}{\mat{Q}}
\newcommand{\Qms}{\mat{Q}_\star}
\newcommand{\Rm}{\mat{R}}
\newcommand{\thetas}{\theta_\star}
\newcommand{\Lamb}{\mat{\Lambda}}
\newcommand{\phiV}{\vec{\phi}}
\newcommand{\psiV}{\vec{\psi}}
\newcommand{\vel}{\vec{u}}
\newcommand{\velt}{\tilde{\vec{u}}}
\newcommand{\grad}{\vec{\nabla}}
\newcommand{\diag}{\mat{diag}}
\renewcommand{\r}{\vec{r}}
\newcommand{\resMorph}{\vec{\mathcal{R}}_\mathsf{M}}

\newcommand{\shearEff}{G_\mathrm{eff}}
\newcommand{\rotMat}[1]{\mat{R}_{#1}}
\newcommand{\projMat}[1]{\mat{\mathcal{P}}_{#1}}
\newcommand{\stress}{\mat{\sigma}}

%% file: acronyms.tex
\begin{acronym}
    \acro{RBC}[RBC]{red blood cell}
\end{acronym}

\begin{acronym}
    \acro{CFD}[CFD]{computational fluid dynamics}
\end{acronym}

\begin{acronym}
    \acro{VAD}[VAD]{ventricular assist device}
\end{acronym}

\begin{acronym}
    \acro{ODE}[ODE]{ordinary differential equation}
\end{acronym}

\begin{acronym}
    \acro{MRF}[MRF]{moving reference frame}
\end{acronym}

\begin{acronym}
    \acro{TTM}[TTM]{tank-treading morphology model}
\end{acronym}

\begin{acronym}
    \acro{TTLM}[TTLM]{logarithmic formulation of the tank-treading morphology model}
\end{acronym}

\begin{acronym}
    \acro{HPC}[HPC]{high performance computing}
\end{acronym}

\begin{acronym}
    \acro{MPI}[MPI]{Message Passing Interface}
\end{acronym}

\begin{acronym}
    \acro{OOP}[OOP]{object-oriented programming}
\end{acronym}

%% file: colors.tex
\definecolor{blau}   {RGB}{  0  84 159}
\definecolor{blau-75}{RGB}{ 64 127 183}
\definecolor{blau-50}{RGB}{142 186 229}
\definecolor{blau-25}{RGB}{199 221 242}
\definecolor{blau-10}{RGB}{232 241 250}

\definecolor{black}   {RGB}{  0   0   0}
\definecolor{black-75}{RGB}{100 101 103}
\definecolor{black-50}{RGB}{156 158 159}
\definecolor{black-25}{RGB}{207 209 210}
\definecolor{black-10}{RGB}{236 237 237}

\definecolor{magenta}   {RGB}{227   0 102}
\definecolor{magenta-75}{RGB}{233  96 136}
\definecolor{magenta-50}{RGB}{241 158 177}
\definecolor{magenta-25}{RGB}{249 210 218}
\definecolor{magenta-10}{RGB}{253 238 240}

\definecolor{yellow}   {RGB}{255 237   0}
\definecolor{yellow-75}{RGB}{255 240  85}
\definecolor{yellow-50}{RGB}{255 245 155}
\definecolor{yellow-25}{RGB}{255 250 209}
\definecolor{yellow-10}{RGB}{255 253 238}

\definecolor{petrol}   {RGB}{  0  97 101}
\definecolor{petrol-75}{RGB}{ 45 127 131}
\definecolor{petrol-50}{RGB}{125 164 167}
\definecolor{petrol-25}{RGB}{191 208 209}
\definecolor{petrol-10}{RGB}{230 236 236}

\definecolor{turkis}   {RGB}{  0 152 161}
\definecolor{turkis-75}{RGB}{  0 177 183}
\definecolor{turkis-50}{RGB}{137 204 207}
\definecolor{turkis-25}{RGB}{202 231 231}
\definecolor{turkis-10}{RGB}{235 246 246}

\definecolor{grun}   {RGB}{ 87 171  39}
\definecolor{grun-75}{RGB}{141 192  96}
\definecolor{grun-50}{RGB}{184 214 152}
\definecolor{grun-25}{RGB}{221 235 206}
\definecolor{grun-10}{RGB}{242 247 236}

\definecolor{maigrun}   {RGB}{189 205   0}
\definecolor{maigrun-75}{RGB}{208 217  92}
\definecolor{maigrun-50}{RGB}{224 230 154}
\definecolor{maigrun-25}{RGB}{240 243 208}
\definecolor{maigrun-10}{RGB}{249 250 237}

\definecolor{orange}   {RGB}{246 168   0}
\definecolor{orange-75}{RGB}{250 190  80}
\definecolor{orange-50}{RGB}{253 212 143}
\definecolor{orange-25}{RGB}{254 234 201}
\definecolor{orange-10}{RGB}{255 247 234}

\definecolor{rot}   {RGB}{204   7  30}
\definecolor{rot-75}{RGB}{216  92  65}
\definecolor{rot-50}{RGB}{230 150 121}
\definecolor{rot-25}{RGB}{243 205 187}
\definecolor{rot-10}{RGB}{250 235 227}

\definecolor{bordeaux}   {RGB}{161  16  53}
\definecolor{bordeaux-75}{RGB}{182  82  86}
\definecolor{bordeaux-50}{RGB}{205 139 135}
\definecolor{bordeaux-25}{RGB}{229 197 192}
\definecolor{bordeaux-10}{RGB}{245 232 229}

\definecolor{violett}   {RGB}{ 97  33  88}
\definecolor{violett-75}{RGB}{131  78 117}
\definecolor{violett-50}{RGB}{168 133 158}
\definecolor{violett-25}{RGB}{210 192 205}
\definecolor{violett-10}{RGB}{237 229 234}

\definecolor{lila}   {RGB}{122 111 172}
\definecolor{lila-75}{RGB}{155 145 193}
\definecolor{lila-50}{RGB}{188 181 215}
\definecolor{lila-25}{RGB}{222 218 235}
\definecolor{lila-10}{RGB}{242 240 247}

%% file: sections_abstract.tex
\begin{abstract}
    The development of blood-handling medical devices, such as ventricular assist devices, requires the analysis of their biocompatibility. Among other aspects, this includes \emph{hemolysis}, i.e., red blood cell damage. For this purpose, \ac{CFD} methods are employed to predict blood flow in prototypes. 
    The most basic hemolysis models directly estimate red blood cell damage from fluid stress in the resulting flow field. More advanced models explicitly resolve cell deformation. On the downside, these models are typically written in a Lagrangian formulation, i.e., they require pathline tracking. We present a new Eulerian description of cell deformation, enabling the evaluation of the solution across the whole domain. The resulting hemolysis model can be applied to any converged \ac{CFD} simulation due to one-way coupling with the fluid velocity field.
    We discuss the efficient numerical treatment of the model equations in a stabilized finite element context. 
    We \replaced{verify}{validate} the model %
    by comparison to the original Lagrangian formulation in selected benchmark flows. Two more complex test cases demonstrate the method's capabilities in real-world applications. The results highlight the advantages over previous hemolysis models. 
    In conclusion, the model holds great potential for the design process of future generations of medical devices.
\end{abstract}
\acresetall

%% file: sections_introduction.tex
\section{Introduction}
\label{sec:introduction}

As the prevalence of heart failure continues to rise worldwide~\cite{vos_global_2020}, the development of blood handling medical devices, such as \acp{VAD}, has become paramount in improving patient outcomes and extending life expectancy. \deleted{Such devices are generally characterized by partly conflicting indicators of performance. \emph{Hydraulic performance}, on the one hand, quantifies the pressure head and the resulting flow rates and encourages high fluid stress. \emph{Hematologic performance}, on the other hand, quantifies the damage to the constituent components of blood and calls for low fluid stress. 
A critical challenge in the development process is thus to balance these aspects.}
Nowadays, the development process is supported by \ac{CFD}. It allows for a comprehensive understanding of how the design and operating parameters of a \ac{VAD} influence the device's \emph{hydraulic performance}. 
Additionally, prototypes need to be assessed for \emph{biocompatibility}. Among other aspects, this involves the quantification of \emph{hemolysis}, i.e., \ac{RBC} damage.
\par
\replaced
{%
In the context of blood-handling medical devices, hemolysis occurs when fluid stresses induce excessive \ac{RBC} deformation, damaging or even rupturing the cell membrane. 
}%
{%
Hemolysis occurs when the hemoglobin contained within an \ac{RBC} leaks out into blood plasma. 
While this can have various causes, we focus on \emph{mechanical} hemolysis in the context of blood-handling medical devices. In mechanical hemolysis, fluid stresses induce excessive cell deformation, which can damage the cell membrane or even lead to cell rupture.%
}
Modeling \deleted{mechanical} hemolysis has been subject of intense research over the past 40 years~\cite{g_heuser_couette_1980,m_giersiepen_estimation_1990,bludszuweit_model_1995,d_arora_tensor-based_2004,pauli_transient_2013,faghih_modeling_2019}. Despite these efforts, there is still no universally accepted approach.
In fact, a recent comparison has shown that computational predictions are still frequently inaccurate~\cite{ponnaluri_results_2023}. This uncertainty, in combination with new findings on the clinical significance of hemolysis~\cite{katz_multicenter_2015}, calls for the development of more reliable hemolysis models.
Existing models can be categorized as \emph{stress-based} models and \emph{strain-based} models. 
\par
Stress-based hemolysis models apply an empirical correlation for he\-mo\-glo\-bin release directly to the instantaneous fluid stress field. This approach has two downsides.
First, such models assume implicitly that cells immediately deform to their steady state when encountering stress. This is because empirical hemolysis correlations are generally measured at constant stress, i.e., the cell has time to adapt to this level of stress and reach a steady state. When encountering changing levels of stress, however, the cell membrane exhibits viscoelastic behavior~\cite{puig_viscoelasticity_2007}. This means that short, high levels of stress may not lead to significant hemolysis if the membrane does not have sufficient time to stretch in response. Stress-based models assume that the full instantaneous fluid stress applies immediately.
Second, it is unclear how the correlation should be applied to arbitrary three-dimensional stress states. %
The empirical correlations~\cite{m_giersiepen_estimation_1990,zhang_study_2011} generally only relate shear stress to hemoglobin release. One generalization to arbitrary stress states was proposed by Bludszuweit~\cite{bludszuweit_model_1995}. She defined a scalar stress that results from a weighted sum of the stress components. 
However, the reduction to a representative scalar loses some information of the three-dimensional stress state, as a cell experiences different loads depending on its alignment with the principal axes of stress. In particular, Goldsmith and Marlow~\cite{goldsmith_flow_1972} showed that cells tend to align more with the flow direction at higher shear rates. 
More recently, it was suggested~\cite{down_significance_2011,faghih_deformation_2020} that for certain flows, extensional stresses have a larger effect on cell deformation than shear stresses. Overall, the proper weighting of the stress components is still unclear and may depend on the flow regimes at hand.
For these reasons, a first-principles model that explicitly describes cell deformation as a result of flow forces is preferable.
\par
Strain-based hemolysis models incorporate such cell deformation models into hemolysis prediction. 
The empirical correlations for hemoglobin release are then applied to the causative membrane strain itself. %
There are different degrees of complexity for the cell deformation model. 
On the simpler side, Chen and Sharp~\cite{chen_strainbased_2011} proposed a scalar model for estimating membrane strain in the context of cell rupture. However, this model was not originally intended for predicting sublethal hemolysis~\cite{chen_testing_2013} and failed to achieve satisfactory results for this purpose~\cite{yu_review_2017}. Arwatz and Smits~\cite{arwatz_viscoelastic_2013} described a scalar viscoelastic model for cell deformation. Its real-world applicability is limited, though, as it does not account for the complex three-dimensional flow forces and it was formulated only for constant shear flow. Thus, simple strain-based models tend to be valid only for specific applications. 
\par
On the more complex side, Ezzeldin et al.~\cite{ezzeldin_strain-based_2015} and Sohrabi and Liu~\cite{sohrabi_cellular_2017} developed models that resolve the cell membrane's three-dimensional deformation. Porcaro and Saeedipour~\cite{porcaro_hemolysis_2023} employed a reduced-order model for the cell structure and modeled the cells' interaction with the flow and with each other. Moreover, there exists a wide array of cell deformation models that were not specifically intended for hemolysis prediction but could be used in this context~\cite{fedosov_multiscale_2010,kloppel_novel_2011,mendez_solver_2014,zavodszky_cellular_2017,kotsalos_bridging_2019,guglietta_effects_2020}. 
These complex approaches
\deleted{resolve cell deformation using varying degrees of freedom and achieve varying levels of computational performance. Nevertheless, they} 
all have in common that their computational cost is too high to simulate large-scale \acp{VAD}. Modern devices support flow rates of up to $10 \, \frac{\mathrm{L}}{\mathrm{min}}$~\cite{foster_third-generation_2018}, corresponding to almost 1 trillion \acp{RBC} per second. This means that \deleted{in these models, generally }only a small fraction of all cells are chosen at the inlet of the device and tracked until the outlet. Hemolysis is then averaged over these cells. This approach has two disadvantages. First, it is unclear how to pick the cells to achieve a representative average. Second, it gives a global index for hemolysis, but does not highlight critical regions inside the domain. In particular, it cannot be guaranteed that the selected cells will penetrate every part of the domain, e.g., boundary layers and recirculation areas. In conclusion, such complex cell deformation models are not well suited to aid the design process of \acp{VAD}. 
\par\bigskip
The aim of the present work is to develop a new model that combines the advantages of the different classes of models described above: the efficiency of stress-based and simple strain-based models, the characteristic cell response to three-dimensional stress of more complex strain-based models, and an Eulerian description, enabling evaluation across the entire domain. 
\par
For this purpose, the strain-based model by Arora et al.~\cite{d_arora_tensor-based_2004} serves as basis, as it provides a compromise between these degrees of complexity, approximating \acp{RBC} as three-dimensional ellipsoids. However, it is based on a Lagrangian description of particle motion. In consequence, its evaluation requires particle tracking and comes with the same drawbacks as the more complex cell deformation models described above. Pauli et al.~\cite{pauli_transient_2013} introduced an Eulerian formulation. In the process, they neglected part of the model to achieve this formulation. As we will show in the present work, this may generally not be an admissible simplification in many situations. Instead, we will derive a full-order Eulerian formulation that is analytically equivalent to the original model. In addition, we will present a modification to improve robustness and efficiency, thus making the model suitable for simulations of realistic \ac{VAD} configurations.
\par
The paper is structured as follows: In \cref{sec:model}, we define the constitutive equations for our strain-based hemolysis model. We introduce the original Lagrangian model formulation and derive the new full-order Eulerian formulation, contrasting it with the previous Eulerian formulation. In addition, we present a novel model with improved efficiency and robustness. In \cref{sec:numerical}, we discuss how we treat the model equations numerically\deleted{ using our in-house multiphysics solver}. In \cref{sec:results}, we show results for \replaced{verification}{validation} and benchmarking. Finally, we summarize our work, discuss the model's limitations and give an outlook on future research in \cref{sec:conclusion}.

%% file: sections_model.tex
\section{Model Equations}
\label{sec:model}

At rest, \aclp{RBC} are known to aggregate into stacks, so-called rouleaux. Under shear, these structures break up and the cells move through the flow in a tumbling motion. At even higher shear rates, cells stop tumbling and start to assume a fixed orientation. In this state, their shape resembles elongated ellipsoids, with their major axis aligned with the flow direction~\cite{h.schmid-schonbeinFluidDropLikeTransition1969,goldsmith_flow_1972}. This motion has been termed tank-treading, as the cell membrane rotates around the cell contents like the treads of a tank~\cite{t.m.fischerRedCellFluid1978}. This regime of high shear rates is of primary interest for hemolysis modeling.
\par
In this section, we present different models to describe \acp{RBC} in this regime. In \cref{sec:model_arora}, we introduce the original Lagrangian model formulation that constitutes the basis for all further discussion. In \cref{sec:model_full-order}, we derive an equivalent Eulerian formulation. In \cref{sec:model_tanktreading}, we present a new model that effectively replicates the original model behavior with better numerical performance. Finally, we contrast this with an older Eulerian model formulation in \cref{sec:model_pauli}. In strain-based hemolysis modeling, any of these \ac{RBC} models may be used to compute a scalar parameter that serves as input to empirical hemolysis correlations. This process is described in \cref{sec:strain-based}.

\subsection{Lagrangian model formulation}
\label{sec:model_arora}

\input{fig_morphology_arora.tex}

Based on the fluid droplet model by Maffettone and Minale~\cite{pl_maffettone_equation_1998}, Arora et al.~\cite{d_arora_tensor-based_2004} derived a Lagrangian cell deformation model. Thereby, \acp{RBC} are assumed as neutrally bouyant ellipsoidal droplets carried with the flow. 
\added{The limitations associated with this assumption are discussed in \cref{sec:conclusion}.}
Ellipsoids can be described mathematically by a symmetric positive definite morphology tensor~$\morph$, whose eigenvalues $\lambda_i$ represent the squared lengths of the ellipsoid's semi-axes. In this work, they are assumed to be in descending order, i.e.,~$\lambda_1 \geq \lambda_2 \geq \lambda_3$. Due to volume conservation, their product has to remain constant. The eigenvectors $\vec v_i$ represent the orientation of the ellipsoid's semi-axes. They can be understood as a rotation matrix $\Qm = [\vec v_1, \vec v_2, \vec v_3]$ that rotates the global inert coordinate system to the local rotating coordinate system of the cell. As derived in \cref{sec:rotation}, the rotation rate of the local frame with respect to the inertial frame is quantified by the rotation tensor 
\begin{equation}
    \label{eq:om}
    \Om = \oddt{\Qm} \Qm\transp \, .
\end{equation}
The cells in the Arora model~\cite{d_arora_tensor-based_2004} do not interact with one another and do not influence the flow field. 
The evolution equation for the morphology of a single cell may be written along its pathline as follows:
\begin{equation}
    \label{eq:model_arora}
    \oddt{\morph} - \comm{\Om}{\morph} = - f_1 \reco{\morph} + f_2 \dist{\strain}{\morph} + f_3 \comm{\vort - \Om}{\morph} \, ,
\end{equation}
with tensor operations $\comm{\Om}{\morph} \coloneqq \Om \morph - \morph \Om$ and $\dist{\strain}{\morph} \coloneqq \strain \morph + \morph \strain$, and coefficients
\begin{equation}
    \label{eq:coefficients}
    f_1 = 5.0 \, \unit{\second}^{-1} \, , \qquad f_2 = f_3 = 4.2298 \cdot 10^{-4} \, .
\end{equation}
For details on the derivation of these coefficients, see~\cite{d_arora_tensor-based_2004}.
The first term on the right-hand side describes shape recovery with the scalar function
\begin{equation*}
    g(\morph) = \frac{6 \, \mathrm{det} \morph}{\mathrm{tr}(\morph)^2-\mathrm{tr}\left( \morph^2 \right)} \, .
\end{equation*}
The second term on the right-hand side describes the effect of fluid strain 
\begin{equation*}
    \strain = \symVelGrad \, ,
\end{equation*}
which causes the cell to align with the principal axes of strain and to deform along those axes. 
The third term on the right-hand side describes the effect of fluid vorticity
\begin{equation*}
    \vort = \antisymVelGrad \, ,
\end{equation*}
which rotates the cell. These three effects are combined as linear superposition.
The left-hand side of \cref{eq:model_arora} represents a Jaumann derivative that defines a co-rotating reference frame. Hence, the model employs a Lagrangian description of cell deformation, so its evaluation requires pathline tracking.  The components of the model are visualized in \cref{fig:morphology_arora}.

\subsection{Full-order Eulerian morphology model}
\label{sec:model_full-order}
A direct reformulation of \cref{eq:model_arora} towards an Eulerian description is challenging due to the implicit dependency between the morphology tensor $\morph$ and the rotation rate of its eigenvectors $\Om$. 
In order to resolve this dependency, we derive an expression for $\Om$ that does not involve derivatives of $\Qm$. For this purpose, the spectral decomposition of the morphology tensor
\begin{equation}
    \morph = \Qm \Lamb \Qm\transp \, , \qquad \Lamb = \diag(\lambda_1, \lambda_2, \lambda_3) \, , \qquad 
    \Qm = [\vec{v}_1, \vec{v}_2, \vec{v}_3] \, ,
    \label{eq:specDecomp}
\end{equation}
is employed. In the following, the model equation~\labelcref{eq:model_arora} is transformed to the eigenbasis of the morphology tensor by multiplying with $\Qm\transp$ from the left and with $\Qm$ from the right. The transformed quantities are then defined as
\begin{equation}
    \label{eq:relative_quantities}
    \straint = \trafoCoord{\strain} \, , \qquad \vortt = \trafoCoord{\vort} \, , \qquad \Omt = \trafoCoord{\Om} \, .
\end{equation}
First, the derivative of the shape tensor $\oddt{\morph}$ becomes
\begin{equation*}
    \label{eq:trafo_matDer}
    \begin{split}
        \Qm\transp \oddt{\morph} \Qm
        &= \Qm\transp \oddt{} \left( \Qm \Lamb \Qm\transp \right) \Qm 
        \\
        &= 
        \Qm\transp 
        \left(
            \oddt{\Qm} \Lamb \Qm\transp
            + \Qm \oddt{\Lamb} \Qm\transp
            + \Qm \Lamb \oddt{\Qm\transp}
        \right)
        \Qm
        \\
        &\overset{\eqref{eq:om}}{=} 
        \Qm\transp \Om \Qm \Lamb
        + \oddt{\Lamb} 
        - \Lamb \Qm\transp \Om \Qm \\
        &= \oddt{\Lamb} + 
        \comm{\Omt}{\Lamb} \, .
    \end{split}
\end{equation*} 
Second, the rotation term $\comm{\Om}{\morph}$ becomes:
\begin{equation*}
    \label{eq:trafo_rot}
    \begin{split}
        \trafoCoord{\comm{\Om}{\morph}} 
        &=
        \Qm\transp \Om \morph \Qm - \Qm\transp \morph \Om \Qm \\
        &= \Qm\transp \Om \Qm \Qm\transp \morph \Qm - \Qm\transp \morph \Qm \Qm\transp \Om \Qm \\
        &= \Omt \Lamb -  \Lamb \Omt \\
        &= \comm{\Omt}{\Lamb} \, .
    \end{split}
\end{equation*}
Third, the same transformation is applied to the right-hand side of \cref{eq:model_arora} in analogous fashion to obtain the full transformed model:
\begin{equation}
    \label{eq:trafo_model}
    \oddt{\Lamb} = -f_1 \reco{\Lamb} + f_2 \dist{\straint}{\Lamb} + f_3 \comm{\vortt - \Omt}{\Lamb} \, .
\end{equation}
Because $\Omt$ and $\vortt$ are antisymmetric and $\Lamb$ is diagonal, the last term on the right-hand side contains no elements on the diagonal. This lets us rewrite the equation only for the off-diagonal elements as follows:
\begin{equation*}
    \comm{\Omt}{\Lamb} = 
    \frac{f_2}{f_3} \dist{\straint}{\Lamb} 
    - \frac{f_2}{f_3} \diag \dist{\straint}{\Lamb}
    + \comm{\vortt}{\Lamb} \, .
\end{equation*}
Here, the $\mat{\diag}$ operator leaves only the diagonal entries and sets all off-diagonal entries to zero.
Applying the inverse coordinate transformation yields: %
\begin{equation}
    \label{eq:om_explicit}
    \comm{\Om}{\morph} = \frac{f_2}{f_3} \dist{\strain - \strainh}{\morph} 
    + \comm{\vort}{\morph} \, , \quad
    \strainh = \Qm \, \diag \left( \Qm\transp \strain \Qm \right) \Qm\transp \, .
\end{equation}
Finally, we obtain the full Eulerian morphology model by treating the morphology tensor $\morph$ as a field variable, hence understanding the Lagrangian derivative as a material derivative. The expression~\labelcref{eq:om_explicit} then allows us to rewrite the original model~\labelcref{eq:model_arora} as follows:
\begin{equation}
    \matDer{\morph} = 
    - f_1 \reco{\morph}
    + f_2 \dist{\strainh}{\morph}
    + \frac{f_2}{f_3} \dist{\strain - \strainh}{\morph}
    + \comm{\vort}{\morph} \, .
    \label{eq:model_full-order}
\end{equation}
This constitutes an analytically equivalent Eulerian formulation of the original model. Intuitively, $\strainh$ represents the component of $\strain$ that solely acts to deform the cell, computed as a projection on the morphology eigenvectors $\Qm$. Consequently, $\strain - \strainh$ represents the component of $\strain$ that solely acts to rotate the cell towards the principal axes of strain. 
The two different strain components are visualized in \cref{fig:morphology_full}.
\par
We remark that there is still an implicit relationship between the morphology tensor $\morph$ and the deformational strain $\strainh$, as the latter is a function of the morphology eigenvectors $\Qm$. In contrast to the original formulation, however, this formulation does not involve the derivatives of these eigenvectors. The dependency is purely algebraic and can be resolved in the framework of the numerical method, e.g., by means of Newton iterations. 
\input{fig_morphology_full.tex}

\subsection{Tank-treading cell deformation model}
\label{sec:model_tanktreading}

An approach to resolve this dependency analytically is to rewrite the model explicitly in terms of the eigenvalues and eigenvectors of $\morph$. 
For this purpose, we consider the diagonal terms and the off-diagonal terms of \cref{eq:trafo_model} separately and use the definitions~\labelcref{eq:om,eq:relative_quantities}:
\begin{subequations} \label{eq:model_eig}
    \begin{align}
        \matDer{\lambda_i} &= -f_1 \left( \lambda_i - g(\Lamb) \right) + 
        2 f_2 \lambda_i \straintI{ii} \, , \label{eq:model_eig_lambda} \\
        \matDer{\Qm} &= \Qm \, \Omt(\Qm) \, , \qquad
        \OmtI{ij}(\Qm) = \frac{f_2}{f_3} \straintI{ij} \frac{\lambda_j + \lambda_i}{ \lambda_j - \lambda_i} +  \vorttI{ij} \, .  \label{eq:model_eig_Q}  
    \end{align}
\end{subequations}
\Cref{eq:model_eig_lambda,eq:model_eig_Q} describe the deformation and rotation of cells, respectively. In particular, the deformation equation~\labelcref{eq:model_eig_lambda} contains the effects of the recovery term (see \cref{fig:morphology_arora_recovery}) and the deformational strain (see \cref{fig:morphology_full_strainDef}). The terms are transformed to the eigensystem of the morphology tensor to act directly on the eigenvalues $\lambda_i$, which represent the ellipsoid's semi-axes. Similarly, the rotation equation~\labelcref{eq:model_eig_Q} describes the effects of the rotational strain (see \cref{fig:morphology_full_strainRot}) and the vorticity (see \cref{fig:morphology_arora_vort}) on the eigenvectors~$\mat Q = [\vec v_1, \vec v_2, \vec v_3]$, which represent the ellipsoid's orientation. 
\par
The rotation equation~\labelcref{eq:model_eig_Q} exhibits a singularity at $\lambda_i = \lambda_j$. This corresponds to a circular cross-section of the ellipsoid, which makes the eigenvectors immediately align with the principal axes of strain in that plane. The eigenvectors may thus be computed as solution to $\straintI{ij} = 0$ in this degenerate state.
\par
The formulation~\labelcref{eq:model_eig} is challenging to solve numerically due to the rotation equation~\labelcref{eq:model_eig_Q}. On the one hand, the rotational source term~\labelcref{eq:model_eig_Q} is three orders of magnitude larger than the deformation source term~\labelcref{eq:model_eig_lambda} and can become arbitrarily large due to the singularity at $\lambda_i = \lambda_j$ This creates a discrepancy of timescales and requires prohibitively small time steps. On the other hand, the orthogonal matrix $\Qm$ is impractical to handle numerically as a field variable, as it contains 9 entries and its columns need to remain orthonormal. These issues lead to large computational effort and a high number of numerical constraints, making direct numerical simulation unfeasible for realistic geometries.
\bigskip\par
To alleviate these issues, we derive a model for the behavior of the rotation equation~\labelcref{eq:model_eig_Q} on the slower timescale of the deformation problem. 
For this purpose, we assume that rotation happens infinitely fast compared to deformation and distinguish between two states: \emph{tank-treading} and \emph{tumbling}. 
\par
\replaced
{
    A \emph{tank-treading} cell is oriented such that the moments of strain and vorticity are in equilibrium, i.e., $\Omt = \zeroMat$ in \cref{eq:model_eig_Q}. The orientation that satisfies this condition is termed $\Qms$ and depends on the cell deformation~$\Lamb$ and the flow quantities~$\strain$ and $\vort$. It varies on the same scale as the flow quantities.
}
{
    A \emph{tank-treading} cell assumes steady orientation with the flow. With the given timescale assumptions, this happens effectively instantly, and the orientation is in equilibrium after the initial alignment process. This essentially corresponds to a quasi-steady state assumption on the orientation in the Lagrangian frame, i.e., $\oddt{\Qm} = \matDer{\Qm} = \zeroMat$. 
    Thus, finding the steady orientation means finding $\Qm$ such that $\Omt = \zeroMat$ in Eq. (9b).
    The $\Qm$ that satisfies this condition is termed $\Qm_\mathrm{steady}$ and depends on the deformation~$\Lamb$ and the flow quantities~$\strain$ and $\vort$.
}
\par
A \emph{tumbling} cell is too deformed to assume \replaced{an equilibrium orientation for the local flow state}{a steady orientation at the given flow state}. In consequence, it will keep rotating and experience tensile and compressional stresses to approximately equal amounts. The source term~$\straint$ in the deformation equation~\labelcref{eq:model_eig_lambda} thus tends to zero in the long timescale. This can be ensured by setting $\Qm = \zeroMat$. 
\par
In sum, the \ac{TTM} becomes:
\begin{subequations}
    \begin{align}
        \matDer{\lambda_i} &= -f_1 \left( \lambda_i - g(\Lamb) \right) + 
        2 f_2 \lambda_i \straintI{ii} \, , \qquad \straint = \trafoCoord{\strain} \, , \label{eq:model_tt_lambda} \\
        \Qm &= 
        \begin{dcases}
            \Qms(\Lamb, \strain, \vort) , & \text{tank-treading} \, , \\
            \zeroMat, & \text{tumbling} \, .
        \end{dcases} \label{eq:model_tt_q}
    \end{align}
    \label{eq:model_tt}
\end{subequations}
\par
This formulation reduces the rotation equation~\labelcref{eq:model_eig_Q} to an algebraic equation~\labelcref{eq:model_tt_q}, avoiding the associated time step limitations. The approach to solve the algebraic equation is detailed in \cref{sec:steady_orientation}. 
Moreover, the deformation equations can be simplified as well; due to volume conservation, the product of the eigenvalues has to remain constant and can be set to unity without loss of generality. Thus, we only need to solve \cref{eq:model_tt_lambda} for $\lambda_1$ and $\lambda_3$ and can compute $\lambda_2 = \frac{1}{\lambda_1 \lambda_3}$. This reduces the number of differential variables from six in the full-order model~\labelcref{eq:model_full-order} to just two in the \ac{TTM}~\labelcref{eq:model_tt}. As demonstrated in \cref{sec:simple_pump}, this results in significant improvements in efficiency and robustness.

\subsubsection{Logarithmic formulation of tank-treading morphology model}

For additional robustness, we employ a logarithmic transformation as presented by Ha\ss{}ler et al.~\cite{hasler_variational_2019} to ensure positive eigenvalues in the correct order. We define the following transformation:
\begin{equation}
    \lambda_1 = 1 + e^{\hat{\lambda}_1} \in (1, \infty)
    \, , \qquad 
    \lambda_3 = \frac{1}{1+e^{\hat{\lambda}_3}} \in (0, 1)
    \, .
    \label{eq:log_trafo}
\end{equation}
We rewrite the material derivative of the deformation model~\labelcref{eq:model_tt_lambda} in terms of the transformed eigenvalues:
\begin{equation}
    \begin{aligned}
        \matDer{\hat{\lambda}_1} &= \frac{\mathcal{F}_1(\Lamb)}{\lambda_1-1}
        \, , \qquad
        \matDer{\hat{\lambda}_3} = \frac{\mathcal{F}_3(\Lamb)}{\lambda_3^2-\lambda_3}
        \, ,
        \\
        \mathcal{F}_i (\Lamb)
        &= -f_1 \left( \lambda_i - g(\Lamb) \right) + 
        2 f_2 \lambda_i \straintI{ii} \, .
    \end{aligned}
    \label{eq:model_ttl}
\end{equation}
This is the \ac{TTLM}.
The model is thus solved for the transformed eigenvalues $\hat{\lambda}_1$ and $\hat{\lambda}_3$. The original eigenvalues $\lambda_1$ and $\lambda_3$ are obtained from the transformation~\labelcref{eq:log_trafo}. The right-hand side source terms are computed using these original eigenvalues.
\input{sections_orientation.tex}

\subsection{Simplified Eulerian formulation}
\label{sec:model_pauli}

Previously, another formulation was obtained by Pauli et al.~\cite{pauli_transient_2013} by neglecting the eigenvector rotation $\Om$ in \cref{eq:model_arora}. This enables a more direct Eulerian description:
\begin{equation}
    \label{eq:model_pauli}
    \matDer{\morph} =  - f_1 \reco{\morph} + f_2 \dist{\strain}{\morph} + f_3 \comm{\vort}{\morph} \, .
\end{equation}
Applying the same transformation as in \cref{sec:model_full-order} yields
\begin{equation}
    \label{eq:om_pauli}
    \comm{\Om}{\morph} = f_2 \dist{\strain - \strainh}{\morph} 
    + f_3 \comm{\vort}{\morph} \, .
\end{equation}
The eigenvector rotation rate $\Om$ is three orders of magnitude smaller than that in the full-order model~\labelcref{eq:om_explicit}. %
The simplified model~\labelcref{eq:model_pauli} is thus not equivalent to the original formulation.
Furthermore, \cref{eq:om_pauli} shows that the simplified model still contains eigenvector rotation, despite initially neglecting the associated terms. These findings indicate that the simplified Eulerian formulation is inconsistent. The effects of this inconsistency will be discussed in \cref{sec:results}.

\subsection{Strain-based hemolysis modeling}
\label{sec:strain-based}

Following Arora et al.~\cite{d_arora_tensor-based_2004}, any of the cell deformation models from above may be employed to compute the cell distortion $D$ and the resulting effective shear rate $\shearEff$:
\begin{equation}
    D = \frac{\sqrt{\lambda_1} - \sqrt{\lambda_3}}{\sqrt{\lambda_1}+\sqrt{\lambda_3}} \, , \qquad
    \shearEff = \frac{2 D f_1}{(1-D^2)f_2} \, .
    \label{eq:shear_eff}
\end{equation}
This effective shear rate may then serve as input to an empirical correlation relating shear stress to hemoglobin release~\cite{m_giersiepen_estimation_1990,zhang_study_2011}. The release rate can be interpreted as source term for an advection-diffusion equation that describes the free hemoglobin concentration in the blood. This has been described in detail in previous works~\cite{goubergrits_numerical_2006,pauli_transient_2013}. The focus of the present study will be the cell morphology model itself, as it constitutes the defining feature of strain-based blood damage models.
\par
\added{It should be noted that despite its name, the effective shear rate $\shearEff$ accounts for both shear and extensional fluid stress, as it is computed directly from the instantaneous cell distortion $D$.
In fact, it provides a natural way to incorporate experimental observations on the significance of extensional stress~\cite{down_significance_2011,faghih_deformation_2020}. In simple shear flow, \acp{RBC} assume a tank-treading orientation that is generally not aligned with the principal axis of strain~\cite{goldsmith_flow_1972}. Thus, the cells experience only a fraction of the total fluid stress. In extensional flow through a contraction, \acp{RBC} align with the principal axis of extension~\cite{faghih_deformation_2020}. As a result, the cells experience the full stress. The predicted distortion $D$ and the resulting effective shear rate $\shearEff$ are thus higher.}

%% file: fig_morphology_arora.tex
\begin{figure}
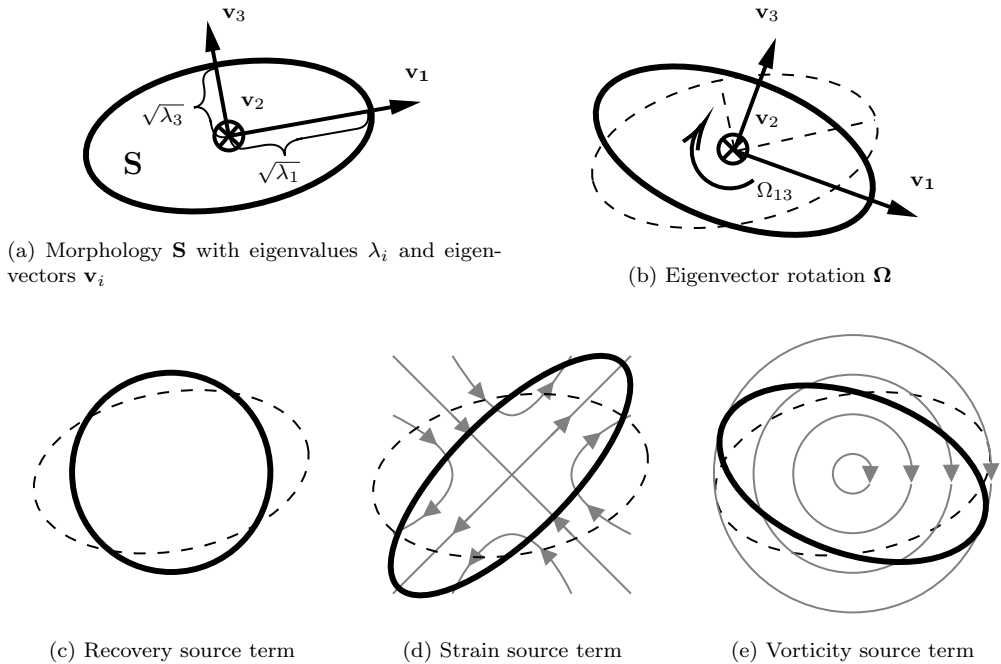

    \begin{subfigure}{0.48\textwidth}
        \centering
        \importpgf{models}{eig_ellipsoid}
        \caption{Morphology $\morph$ with eigenvalues $\lambda_i$ and eigenvectors $\vec{v}_i$}
        \label{fig:morphology_arora_ellipsoid}
    \end{subfigure}
    \begin{subfigure}{0.48\textwidth}
        \centering
        \importpgf{models}{eig_rotation}
        \caption{Eigenvector rotation $\Om$}
        \label{fig:morphology_arora_rotation}
    \end{subfigure}
    \\[1em]
    \begin{subfigure}{0.32\textwidth}
        \centering
        \importpgf{models}{source_recovery}
        \caption{Recovery source term}
        \label{fig:morphology_arora_recovery}
    \end{subfigure}
    \begin{subfigure}{0.32\textwidth}
        \centering
        \importpgf{models}{source_strain}
        \caption{Strain source term}
        \label{fig:morphology_arora_strain}
    \end{subfigure}
    \begin{subfigure}{0.32\textwidth}
        \centering
        \importpgf{models}{source_vort}
        \caption{Vorticity source term}
        \label{fig:morphology_arora_vort}
    \end{subfigure}
    \caption{Ellipsoidal cell representation in Arora morphology model}
    \label{fig:morphology_arora}
\end{figure}

%% file: fig_morphology_full.tex
\begin{figure}
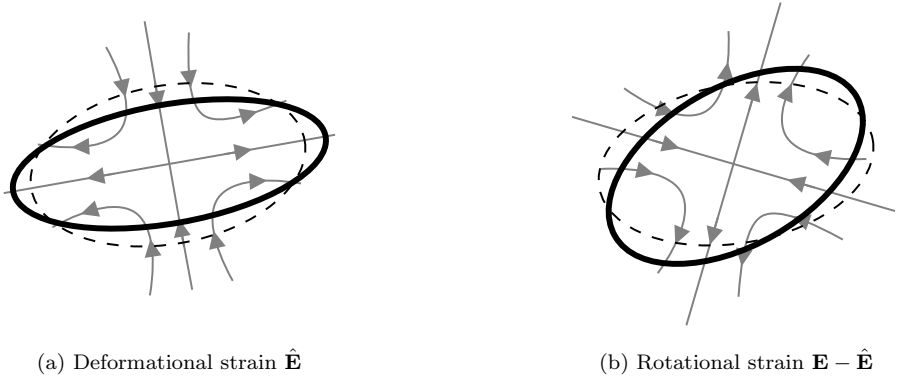

    \centering
    \begin{subfigure}{0.45\textwidth}
        \centering
        \importpgf{models}{strain_def}
        \caption{Deformational strain $\strainh$}
        \label{fig:morphology_full_strainDef}
    \end{subfigure}
    \hfill
    \begin{subfigure}{0.45\textwidth}
        \centering
        \importpgf{models}{strain_rot}
        \caption{Rotational strain $\strain - \strainh$}
        \label{fig:morphology_full_strainRot}
    \end{subfigure}
    \caption{Strain terms in full-order Eulerian formulation}
    \label{fig:morphology_full}
\end{figure}

%% file: sections_orientation.tex
\subsubsection{Determining the \replaced{equilibrium}{steady} orientation of the cell}
\label{sec:steady_orientation}
The algebraic equation~\labelcref{eq:model_tt_q} requires us to solve the equation \deleted{$\Om(\Qm) = \zeroMat$ or, equivalently,} $\Omt(\Qm_\star) = \zeroMat$.
We will derive the solution analytically in 2D first and then generalize it numerically to 3D.
\replaced{In 2D, we assume without loss of generality that any rotation happens in the plane that contains the ellipsoid's first two axes $\vec{v}_1, \vec{v}_2$}
{Without loss of generality, assume that the two-dimensional rotation happens in the plane normal to the third axis} and that $\lambda_1 \ne \lambda_2$ (otherwise the cross-section is circular and the orientation is arbitrary). Then we only need to consider $\strain , \vort, \Lamb, \Qm, \Om \in \mathbb{R}^{2\times2}$. In particular, the two-dimensional orientation $\Qm = [\vec{v}_1, \vec{v}_2]$ can be expressed \replaced{as a rotation~$\Rm$ of the reference orientation $\unitM$ by an angle~$\theta$:}
{in terms of a single variable~$\theta$:}
\begin{equation}
    \Qm  = \added{\Rm(\theta) =}
    \begin{bmatrix}
        \cos \theta & -\sin \theta \\
        \sin \theta & \cos \theta \\
    \end{bmatrix} \, . 
    \label{eq:rotation_matrix}
\end{equation}
This allows us to write the rotation rate~\labelcref{eq:model_eig_Q} as a function of this scalar $\theta$:
\begin{equation}
    \OmtI{21}(\theta; \Lamb, \strain, \vort) = 
    \frac{f_2}{f_3} \frac{\lambda_1 + \lambda_2}{\lambda_1 - \lambda_2} \left( \strainI{12} \cos(2 \theta) + \frac{1}{2} (\strainI{22}-\strainI{11}) \sin(2 \theta) \right) - \vortI{12} \, .
    \label{eq:rotation_rate_2d}
\end{equation}
For ease of notation, we define the coefficients
\begin{equation}
    \begin{split}
    a(\Lamb, \strain) &\coloneqq \frac{f_2}{f_3} \frac{\lambda_1 + \lambda_2}{\lambda_1 - \lambda_2} \strainI{12} \, , \qquad
    b(\Lamb, \strain) \coloneqq \frac{f_2}{2 f_3}  \frac{\lambda_1 + \lambda_2}{\lambda_1 - \lambda_2} (\strainI{22}-\strainI{11}) \, , \\
    c(\vort) &\coloneqq \vortI{12} \, .
    \end{split}
    \label{eq:orientation_coefficients}
\end{equation}
\replaced{
The equilibrium orientation angle $\thetas$ is a solution to the equation $\OmtI{21}(\thetas) = 0$. Mathematical analysis of \cref{eq:rotation_rate_2d} reveals that such a solution exists only if}
{A solution to $\OmtI{21} = 0$ exists only if}
\begin{equation}
    a^2 + b^2 \ge c^2 \, .
    \label{eq:discr}
\end{equation}
This condition weighs the aligning effects of strain against the rotational effects of vorticity. If the vorticity is too strong, the strain cannot keep the cell \replaced{at equilibrium}{at a fixed orientation} and the cell tumbles instead. If \replaced{an equilibrium}{a steady} orientation exists, there are always two solutions. We demand that the \replaced{equilibrium}{steady orientation} is stable, i.e., the cell returns to \replaced{equilibrium}{the steady orientation} after small perturbations. This corresponds to the condition $\pd{\OmtI{21}}{\theta} < 0$. \replaced{Analysis of \cref{eq:rotation_rate_2d} shows that}{We find that} the only solution that satisfies this condition is
\begin{equation}
    \thetas(a, b, c) = \arctan 
    \left( 
        \frac{a + \sqrt{a^2+b^2 - c^2}}{b+c}
    \right) \, .
    \label{eq:theta_steady}
\end{equation}
This corresponds to the solution where the longest semi-axis of the cell experiences \deleted{the most} tension and the shortest semi-axis experiences \deleted{the most} compression.
The \replaced{equilibrium}{steady} orientation may then be computed \replaced{from \cref{eq:rotation_matrix} as $\Qms = \Rm(\thetas)$.}{as $\Qms = \Qm(\thetas)$.} 
\added{This is illustrated in \cref{fig:equilibrium_orientation_2d}.}
\input{fig_equilibrium_orientation.tex}
\bigskip\par
Next, we will present an approach to apply this analytical \added{two-dimensional} solution in a three-dimensional setting, i.e., $\strain , \vort, \Lamb, \Qm, \Om \in \mathbb{R}^{3\times3}$. 
For this purpose, we define the \emph{projection} operator
\begin{equation*}
    \projMat{k} : \mathbb{R}^{3 \times 3} \to \mathbb{R}^{2 \times 2} \, , \quad \projMat{k}(\mat A) = (A_{ij})_{i \ne k, \\ j \ne k} \, .
\end{equation*}
It projects a three-dimensional state to the plane normal to the $k$-th coordinate axis \added{by deleting the $k$-th row and the $k$-th column of the tensor $\mat{A}$. For each axis $k$, we obtain two-dimensional flow quantities $\strain^{(k)} = \projMat{k}(\strain), \ \vort^{(k)} = \projMat{k}(\vort)$ and a two-dimensional cell deformation $\Lamb^{(k)} = \projMat{k}(\Lamb)$ for the respective cross-section of the ellipsoid. We can then treat each cross-section according to the two-dimensional findings from above. This allows us to determine the equilibrium angle $\theta_k$ in that cross-section $k$. This is illustrated in \cref{fig:equilibrium_orientation_3d}.}
\begin{algorithm}
    \SetAlgoLined
    \KwIn{A cell deformation $\Lamb$, fluid strain $\strain$ and fluid vorticity $\vort$}
    \KwOut{A cell orientation $\Qm$ for the \ac{TTM} within an accuracy of $\varepsilon$.}
    \SetKwRepeat{Do}{do}{while}
    $\Qm \gets \unitM$  \;
    \Do{not converged}{ 
        $converged \gets true$ \;
        \For{$k\gets1$ \KwTo $3$ }
        {
            $\straint \gets \trafoCoord{\strain} \, , \quad \vortt \gets \trafoCoord{\vort}$
            \label{line:transformations}%
            \;
            $\strain^{(k)} \gets \projMat{k}(\straint) \, , \quad
            \vort^{(k)} \gets \projMat{k}(\vortt) \, , \quad
            \Lamb^{(k)} \gets \projMat{k}(\Lamb) $ 
            \label{line:projections}%
            \;
            $a_k \gets a(\Lamb^{(k)}, \strain^{(k)}) \, , \
            b_k \gets b(\Lamb^{(k)}, \strain^{(k)}) \, , \
            c_k \gets c(\vort^{(k)})$ 
            \label{line:coefficients}%
            \;
            \uIf(\tcp*[f]{cell is tank-treading}){$a_k^2 + b_k^2 \ge c_k^2$}
            {
                $\theta_k \gets \thetas(a_k, b_k, c_k)$ 
                \label{li:theta}%
                \;
                $\Qm \gets \Qm \rotMat{k}(\theta_k)$ \label{li:underrelaxation} \; 
                $converged \gets (converged \land |\theta_k| \le \varepsilon)$ \;
            }
            \Else(\tcp*[f]{cell is tumbling}){
                $\Qm \gets \zeroMat$ 
                \label{li:tumbling}
                \;
                return \;
            }
        }
    }
    \caption{\replaced{Equilibrium}{Steady} orientation algorithm in 3D}
    \label{alg:steady}
\end{algorithm}
\par
\added{All required steps are listed in Algorithm~\ref{alg:steady}. We iteratively approximate the equilibrium orientation $\Qm_\star$ by successively applying rotations $\theta_k$ around the ellipsoid's axes $\vec{v}_k$. The current approximation is stored in $\Qm = [\vec{v}_1, \, \vec{v}_2, \, \vec{v}_3]$. Each iteration works as follows: First, we use the current orientation $\Qm$ to transform the fluid strain and vorticity tensors to the coordinate system of the ellipsoid according to \cref{eq:relative_quantities} (see line~\ref{line:transformations}). The transformation ensures that
the subsequent projection (see line~\ref{line:projections}) gives us information in a plane that contains two of the ellipsoid's axes $\vec v_i , \, i \neq k$. This is important because the deformation $\Lamb^{(k)} = \projMat{k}(\Lamb)$ is given in this plane. Next, we treat this cross-section according to the two-dimensional results; we compute the coefficients~$a_k, \, b_k, \, c_k$ using \cref{eq:orientation_coefficients}. Based on the condition~\labelcref{eq:discr}, we determine if an equilibrium orientation exists. If it does, we compute the equilibrium angle $\theta_k$ according to \cref{eq:theta_steady} (see line~\ref{li:theta}). Analogous to \cref{eq:rotation_matrix}, we define an elementary rotation matrix $\Rm_k(\theta) \in \mathbb{R}^{3x3}$ to perform a rotation around the $k$-th axis. We apply the rotation to the current orientation $\Qm$ by multiplying from the right (see line~\ref{li:underrelaxation}). This is because we want to rotate around the $k$-th axis of the ellipsoid $\vec{v}_k$, rather than the $k$-th axis of the reference coordinate system. If the equilibrium orientation does not exist, the cell is tumbling. According to \cref{eq:model_tt_q}, we set $\Qm = \zeroMat$ (see line~\ref{li:tumbling}).}
\par
In case of unstable convergence behavior, an underrelaxation factor may be added to the update of the orientation in line~\ref{li:underrelaxation}. In the underlying study, Algorithm~\labelcref{alg:steady} converged for over $99.9 \, \%$ of quadrature points, %
mostly within fewer than 10 iterations of the outer loop. The algorithm converges towards a state where the longest axis of the cell experiences \deleted{the most} tension and the shortest axis experiences \deleted{the most} compression, i.e., $\straintI{11} \ge \straintI{22} \ge \straintI{33}$. This is in agreement with the two-dimensional results.
Overall, we find that Algorithm~\ref{alg:steady} reliably and efficiently produces a unique solution.
\par
An alternative is numerically integrating the transient equation $\od{\Qm}{t} = \Qm \Omt$ up until steady state. This is computationally inefficient, as it requires on the order of 100,000 time steps and does not allow for the detection of tumbling. It is used only as a fallback solution in case Algorithm~\labelcref{alg:steady} does not converge.

%% file: fig_equilibrium_orientation.tex
\begin{figure}
    \centering
    \begin{subfigure}{0.4\textwidth}
        \centering
        {\small ${\color{blau-75}\strain}, \, {\color{grun} \vort}, \, {\color{bordeaux} \Lamb}, \, \Qm \in \mathbb{R}^{2\times2}$}
        \importpgf{models}{ell-steady}
        \caption{Equilibrium orientation angle in 2D}
        \label{fig:equilibrium_orientation_2d}
    \end{subfigure}
    \begin{subfigure}{0.4\textwidth}
        \centering
        \begin{overpic}[width=\textwidth]{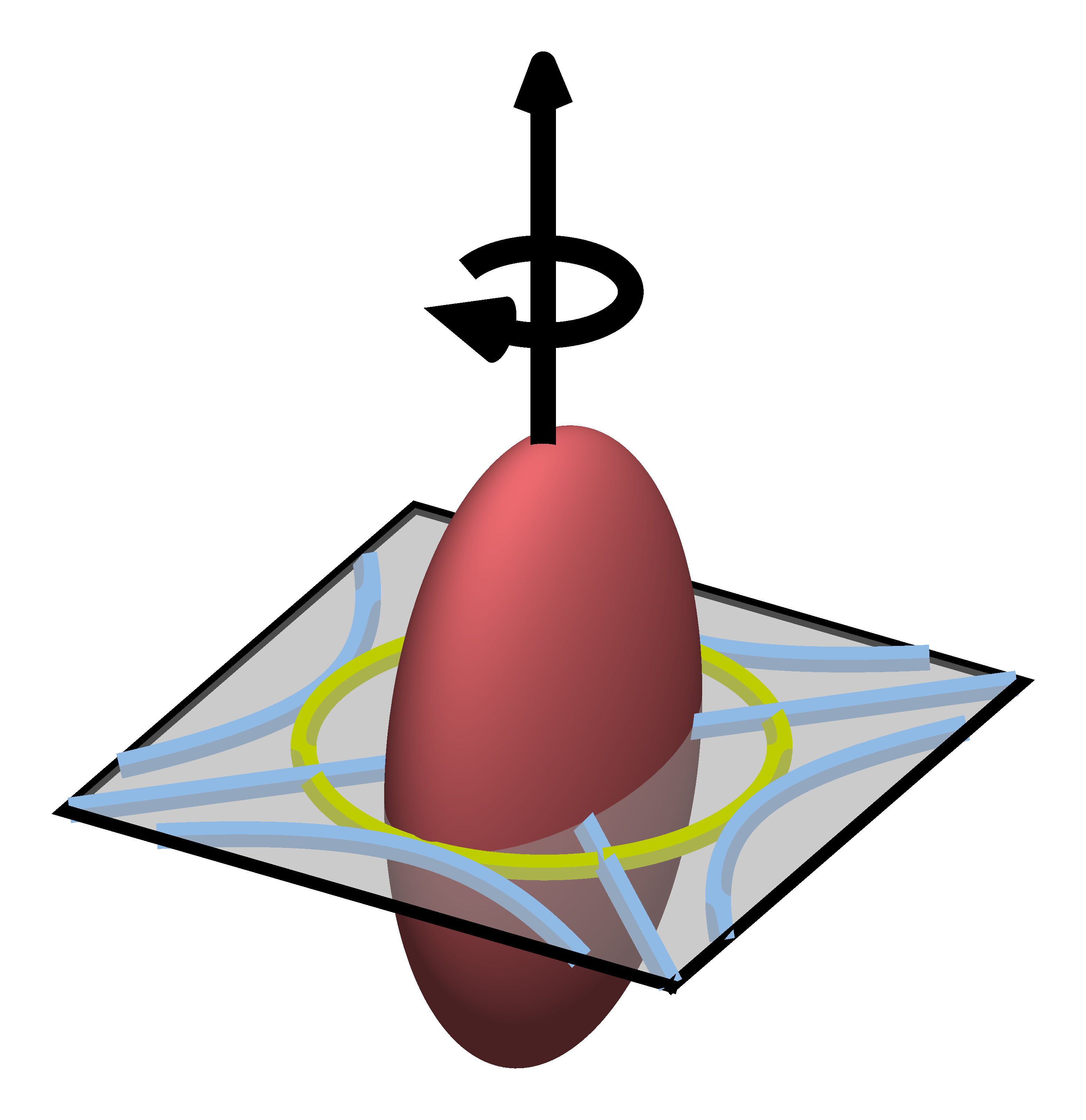}
            \put (54,95) {\small \color{black} $\vec v_k$}
            \put (65,50) {\small ${\color{blau-75}\strain^{(k)}}, \, {\color{grun} \vort^{(k)}}, \, {\color{bordeaux} \Lamb^{(k)}}$}
            \put (95,40) {\small $\in \mathbb{R}^{2\times2}$}
            \put (30,75) {\small $\theta_k$}
        \end{overpic}
        \caption{Projection to $k$-th normal plane in 3D}
        \label{fig:equilibrium_orientation_3d}
    \end{subfigure}
    \caption{Determining equilibrium orientation in 2D and 3D}
    \label{fig:equilibrium_orientation}
\end{figure}
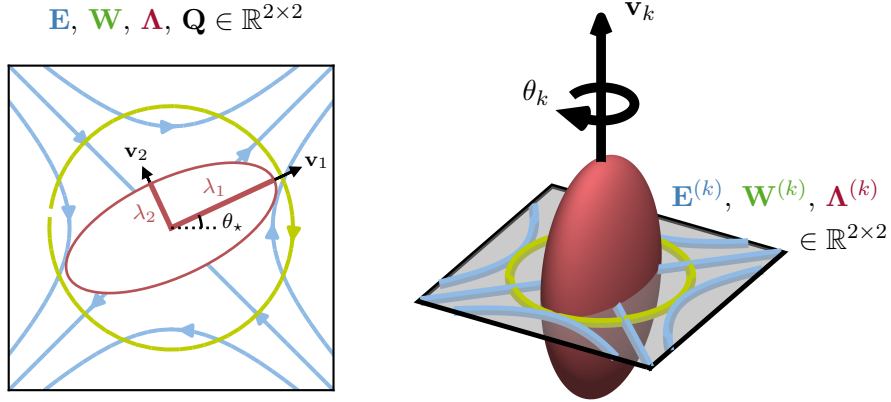

%% file: sections_numerical.tex
\section{Numerical Method}
\label{sec:numerical}

Our simulations consist of two components: the flow problem and the morphology problem. The flow problem is solved first and information about the flow field is then transferred to the morphology problem in a one-way coupling. 
In this section, we first present the mathematical definitions of these two problems in \cref{sec:flow_problem,sec:morphology_problem}. The computational frameworks that implement these problems are described in \cref{sec:lagrangian_verification}.

\subsection{Flow problem}
\label{sec:flow_problem}

Whole blood is assumed to be an incompressible, Newtonian fluid, which is generally valid for blood at high shear rates~\cite{pauli_stabilized_2016}. High shear regions are the most significant in this context, since this is where the majority of hemolysis occurs. Fluid flow is thus governed by the incompressible Navier-Stokes equations:
\begin{align*}
    \rho 
    \left(
        \pd{\vel}{t}
        + (\vel \cdot \grad) \vel
    \right)
    &= - \grad \cdot \stress 
    && \text{in $\domain$}
    \, , \\
    \grad \cdot \vel &= 0 
    && \text{in $\domain$}
    \, , \\
    \stress &= -p \unitM + 2 \mu \strain
    && \text{in $\domain$}
    \, , \\
    \vel &= \vel_\mathrm{in}
    && \text{on $\Gamma_\mathrm{in}$}
    \, , \\
    \vel &= \vel_\mathrm{wall}
    && \text{on $\Gamma_\mathrm{wall}$}
    \, , \\
    \stress \cdot \vec{n} &= \zeroVec
    && \text{on $\Gamma_\mathrm{out}$}
    \, ,
\end{align*}
for the fluid domain $\domain$. The domain boundary $\partial \domain$ is partitioned into inflow $\Gamma_\mathrm{in}$, no-slip walls $\Gamma_\mathrm{wall}$ and stress-free outflow $\Gamma_\mathrm{out}$.
As material parameters for blood, we select a density of $\rho = 1.054 \, \unit{\gram\per\cubic\centi\metre}$ and a viscosity of $3.5 \, \mathrm{cP}$. 
The problem is discretized with stabilized finite elements.
For details on the stabilization terms, we refer to Pauli and Behr~\cite{pauli_stabilized_2017}.
To simulate rotating systems at steady state, we employ the \ac{MRF} approach~\cite{pauli_stabilized_2015}. \added{For our applications involving an impeller, we define a rotating inner domain $\domainr$ and a fixed outer domain $\domain \setminus \domainr$. The inner domain rotates along with the impeller at an angular velocity $\Omr$.}

\subsection{Morphology problem}
\label{sec:morphology_problem}

Fundamentally, the morphology models~\labelcref{eq:model_full-order,eq:model_tt,eq:model_pauli} represent ad\-vec\-tion-reaction equations. %
Aggregating the respective differential degrees of freedom in a vector $\phiV$, each of them can be written in the following form:
\begin{subequations}
    \begin{alignat}{2}
        \resMorph(\phiV; \vel, \grad\vel) \coloneqq 
        \matDer{\phiV} 
        - \vec{f} (\phiV, \grad\vel)
        &= \zeroVec 
        \qquad && \text{in $\domain$} \, , 
        \label{eq:morphology_models_residual} \\
        \phiV &= \phiV_\mathrm{in}  %
        \qquad && \text{on $\Gamma_\mathrm{in}$} \, , \\
        \grad \phiV \cdot \vec{n} &= \zeroVec
        \qquad && 
        \text{on $\Gamma_\mathrm{wall} \cup \Gamma_\mathrm{out}$} \, .
    \end{alignat}
    \label{eq:morphology_models_general}
\end{subequations}
At the inlet $\Gamma_\mathrm{in}$, we impose some cell morphology, e.g., an undeformed cell. 
At walls and at the outflow, there is no component of advection velocity from the boundary into the computational domain. Consequently, we apply no-flux boundary conditions at all remaining boundaries $\Gamma_\mathrm{wall} \cup \Gamma_\mathrm{out} = \partial \domain \setminus \Gamma_{\mathrm{in}}$. 

\added{Using the MRF approach, the governing equations \labelcref{eq:morphology_models_residual} are transformed to the same reference frames as the flow problem. In particular, \acp{RBC} are advected by the relative velocity field $\velt = \vel - \Omr \times \r$ in the rotating frame. The flow forces acting on the cells, i.e., the velocity gradients, are transformed accordingly:}
\begin{equation*}
    \resMorph(\phiV; \velt, \grad\vel - \Omr)
    = \zeroVec
    \qquad \text{in $\domainr$} \, .
\end{equation*}
This formulation holds true for model formulations that are \emph{objective}, i.e., invariant under rotation. This is the case for models~\labelcref{eq:model_full-order,eq:model_tt}. The simplified model~\labelcref{eq:model_pauli} is not objective, however, as cells do not rotate with the full fluid vorticity $\vort$. This particular model thus requires the addition of another term that accounts for the rotating reference frame:
\begin{equation*}
    \pd{\morph}{t} 
    + (\velt \cdot \grad) \morph
    = 
    - f_1 \reco{\morph} 
    + f_2 \dist{\strain}{\morph} 
    + f_3 \comm{\vort}{\morph} 
    - \comm{\Omr}{\morph} 
    \qquad \text{in $\domainr$} \, .
\end{equation*}

For the finite element discretization, we consider the stabilized weak Galerkin formulation of the problem: Find $\phiV^h \in \mathcal{S}^h \subset [H^1(\domain)]^{n_\mathrm{dof}}$ such that:
\begin{equation}
    D_\mathsf{M}(\psiV^h, \phiV^h) 
    + E_\mathsf{M}(\psiV^h, \phiV^h) 
    + J_\mathsf{M}(\psiV^h, \phiV^h) 
    = 0 \qquad 
    \forall \psiV^h \in \mathcal{V}^h \subset [H^1_0(\domain)]^{n_\mathrm{dof}} \, ,
    \label{eq:morphology_weak_form}
\end{equation}
where $H^1(\domain)$ denotes the set of functions with square-integrable first derivatives on $\domain$ and $H^1_0(\domain)$ is a subset of those functions satisfying a homogeneous Dirichlet boundary condition on $\Gamma_\mathrm{in}$. 
The first term represents the constitutive model equations:
\begin{equation*}
    \begin{split}
    D_\mathsf{M}(\psiV^h, \phiV^h) 
    = & 
    \int_{\domain \setminus \domainr} \psiV^h \cdot
    \resMorph(\phiV^h, \vel, \grad\vel)
    \, \mathrm d\intMeasure \\
    + &
    \int_{\domainr} \psiV^h \cdot
    \resMorph(\phiV^h, \velt, \grad\vel - \Omr)
    \, \mathrm d\intMeasure \, . 
    \end{split}
\end{equation*}
The second term $E_\mathsf{M}$ represents stabilization. We are using Galerkin/least-squares (GLS) stabilization~\cite{tjr_hughes_new_1986}. For details on the formulation for multi-dimensional variables, we refer to Pauli~\cite{pauli_stabilized_2016}.
The third term represents $YZ\beta$ discontinuity capturing~\cite{bazilevs_yz_2007}, computed element by element:
\begin{equation*}
    J_\mathsf{M}(\psiV^h, \phiV^h) =
    \sum_{e=1}^{n_\mathrm{el}} 
    \int_{\domain_e} 
    \nu_\mathrm{DC} 
    \grad \psiV^h : \grad \phiV^h
    \, \mathrm d\intMeasure \, .
\end{equation*}

\subsection{Implementation}
\label{sec:lagrangian_verification}

\added{
    The Eulerian problems presented in \cref{sec:flow_problem,sec:morphology_problem} are solved using our in-house multiphysics finite element code \texttt{XNS}. Details on the implementation can be found in \cref{sec:implementation}.
}

\input{sections_lagrangian.tex}

%% file: sections_lagrangian.tex
In order to \replaced{verify}{validate} the Eulerian model formulations and their numerical discretization, we compare them to their Lagrangian counterparts. This is achieved in five steps, all of which are available as part of our open-source Python package HemTracer\footnote{\url{https://github.com/nicodirkes/HemTracer}}.
First, we use the same Eulerian flow field determined by the flow problem (see \cref{sec:flow_problem}) to compute pathlines, i.e., trajectories of fluid particles. They are obtained by integrating the following \ac{ODE} starting from a given initial position~$\vec{X}_0$:
\begin{equation}
    \label{eq:pathline}
    \oddt{\vec X} = 
    \begin{dcases}
        \vel(\vec{X}, t) \, , & \text{$\vec{X}(t) \in \domain \setminus \domainr$} \, , \\
        \velt(\vec{X}, t) \, , & \text{$\vec{X}(t) \in \domainr$} \, ,
    \end{dcases}
    \qquad
    \vec{X}(0) = \vec{X}_0 \, ,
\end{equation}
Second, we compute the velocity gradients of the flow field and interpolate them to the pathlines $\vec{X}(t)$ to obtain $\grad\vel(t)$. Third, we rewrite the morphology problem in the Lagrangian frame by substituting the derivatives with respect to space and time with the material derivatives along the pathline, i.e., the morphology problem \labelcref{eq:morphology_models_general} becomes:
\begin{equation*}
    \oddt{\phiV} 
    = 
    \begin{dcases}
        \vec{f} (\phiV, \grad\vel(t)) \, , & \text{$\vec{X}(t) \in \domain \setminus \domainr$} \, , \\
        \vec{f} (\phiV, \grad\vel(t) - \Omr) \, , & \text{$\vec{X}(t) \in \domainr$} \, ,
    \end{dcases}
    \qquad
    \phiV(0) = \phiV_0 \, .
\end{equation*}
We match the initial condition $\phiV_0$ to the Eulerian morphology field at the initial position $\vec{X}_0$.
Fourth, we solve the Lagrangian problem numerically by integrating the above \ac{ODE} in time. We use the \texttt{scipy.integrate.solve\_ivp} routine with an adaptive Runge-Kutta 45 scheme.
Fifth, we interpolate the Eulerian morphology results to the same pathline and compare them to the Lagrangian results.

%% file: sections_results.tex
\section{Numerical Results}
\label{sec:results}
In order to compare the models presented above, we apply them to a selection of benchmark problems. In \cref{sec:simple_shear,sec:rotating_shear}, we consider two simple flows to \replaced{verify}{validate} the new Eulerian model formulations~\labelcref{eq:model_full-order,eq:model_tt} and show the deficiencies of the simplified model~\labelcref{eq:model_pauli}. \added{We discuss two more realistic configurations in \cref{sec:pauli_stir,sec:simple_pump} to show practical advantages of the \acf{TTM}~\labelcref{eq:model_tt} over the full-order model~\labelcref{eq:model_full-order}.}
\subsection{Simple shear}%
\label{sec:simple_shear}%
\input{fig_simpleShear.tex}First, we consider planar two-dimensional Couette flow. 
This configuration %
is visualized in \cref{fig:simpShear_setup}. The channel height is set to $h = 2.5 \cdot 10^{-5} \, \unit{\metre}$ and the top wall moves at $U = 1 \, \unit{\metre\per\second}$. The channel has a streamwise length of $2 \, \unit{\metre}$.
The computational domain is discretized with $10{,}000$ rectangular finite elements. At the inlet, the elements have a streamwise length of $5\cdot 10^{-6} \, \unit{\metre}$ to capture cell behavior in the entry region. The length progressively increases downstream.
At the inlet, red blood cells are introduced with an initial shape of $(\lambda_1, \lambda_2, \lambda_3) = (2, 1, 0.5)$. 
The angle between the flow direction and the major axis is defined as $\theta$. 
At the inlet, the orientation is imposed such that the major axis is orthogonal to the flow direction, i.e., $\theta_\mathrm{in} = 90 \unit{\degree}$. The solutions for the Eulerian models~\labelcref{eq:model_full-order,eq:model_tt,eq:model_pauli} are obtained by solving the steady state morphology problem~\labelcref{eq:morphology_weak_form}.
\par
From a Lagrangian point-of-view, planar Couette flow produces simple shear, i.e., unidirectional flow with a constant gradient perpendicular to the flow direction. Fluid strain and vorticity are of equal strength in simple shear. Their magnitude is defined by the fluid shear rate $G_\mathrm{f} = 40{,}000 \, \unit{\per\second}$. If we align the $x$-axis with the flow direction and the $y$-axis with the gradient direction, the flow gradients can be written as follows:
\begin{equation*}
    \grad\vel = 
    \begin{bmatrix}
        0 & G_\mathrm{f} & 0\\
        0 & 0 & 0\\
        0 & 0 & 0
    \end{bmatrix} 
    \, . 
\end{equation*}
For this flow field, pathlines are straight lines with $y = \mathrm{const}$. We evaluate the Lagrangian Arora model along these pathlines as described in \cref{sec:lagrangian_verification}. The initial condition is equivalent to the inlet boundary condition then. If we choose the pathline at the top, i.e., $y = h$ with $U = 1 \, \unit{\metre\per\second}$, the time coordinate $t$ in the Lagrangian formulation corresponds one-to-one to the spatial coordinate $x$ in the Eulerian formulation. The solution of the Lagrangian model can thus be compared directly to the Eulerian solution.
\par
\Cref{fig:simpShear_phi} shows the evolution of the major axis angle $\theta$. Starting from the imposed initial angle of $90^\circ$, the cell rapidly assumes an orientation~$\theta = 26.6^\circ$. As visualized in the inset in \cref{fig:simpShear_phi}, the initial alignment happens within the first $0.1 \, \unit{\milli\second}$ according to the Arora model~\labelcref{eq:model_arora}. The resulting angle corresponds precisely to the \replaced{equilibrium}{steady} orientation for the initial deformation and the given flow state, i.e., $\thetas = 26.6^\circ$ according to \cref{eq:theta_steady}. Downstream, the cell remains \replaced{at equilibrium}{at steady orientation} orientation $\thetas$, which decreases asymptotically towards 0 with increasing cell deformation. This steady orientation represents a tank-treading state. The alignment of the major axis with the flow direction at high shear rates agrees with experimental observations~\cite{h.schmid-schonbeinFluidDropLikeTransition1969,goldsmith_flow_1972}.
\par
The Eulerian models differ in their ability to predict the initial rapid alignment. As shown in the inset, the full-order model~\labelcref{eq:model_full-order} predicts the alignment perfectly. The \ac{TTM} does not resolve the initial alignment, as the cell is assumed to be at \replaced{equilibrium}{steady state} permanently. However, it predicts the orientation accurately after the initial alignment, i.e., $x > 1 \cdot 10^{-4} \, \unit{\metre}$. In contrast, the simplified model~\labelcref{eq:model_pauli} predicts a slower alignment with the tank-treading orientation. This is due to the lower rotation rate of cells (cf. \cref{sec:model_pauli}). The final orientation is predicted accurately by the simplified model. This is because \replaced{it}{the steady orientation} represents an equilibrium between the rotational effects of strain and vorticity. The strength of these effects relative to one another is the same in the simplified model as in the full-order model. The absolute rotation rate is lower, however, causing slower transient behavior. 
\par
\Cref{fig:simpShear_l1} shows the evolution of the major axis length $\sqrt{\lambda_1}$. Starting from the initial deformation, the cell deforms towards a steady state deformation of $(\lambda_1, \lambda_2, \lambda_3) = (10.5, 0.43, 0.22)$. According to \cref{eq:shear_eff}, this corresponds to an effective shear rate of $G_\mathrm{eff} = 40{,}000 \, \unit{\per\second} = G_\mathrm{f}$. This is expected behavior, as the effective shear rate is constructed precisely to predict an equivalent simple shear flow that induces the instantaneous cell deformation at steady state.
The Eulerian models agree in the prediction of this final deformation, but they differ in their prediction of transient behavior.
The full-order model predicts the evolution of deformation in full agreement with the Arora model. The \ac{TTM} provides an equivalent approximation. The discrepancy with respect to the initial cell orientation is limited to such small timescales that it does not affect deformation, which occurs over longer timescales. The simplified model, on the other hand, significantly overpredicts intermediate deformation. This is due to the slower alignment with the tank-treading state, which causes the cell to be more aligned with the direction of shear for longer. The steady deformation is predicted correctly due to the correct prediction of the final orientation. 
\bigskip\par
\input{fig_simpleShear_shortExposure.tex}
\added{
To investigate the behavior under shorter exposure times, we modify the configuration in \cref{fig:simpShear_setup} by increasing the top wall velocity to $U = 200 \, \unit{\metre\per\second}$ and increasing the bottom wall velocity to $U_\mathrm{bot} = 100 \, \unit{\metre\per\second}$. With a gap width of $h = 1 \cdot 10^{-4} \, \unit{\metre}$, this results in a fluid shear rate of $G_\mathrm{f} = 1{,}000{,}000 \, \unit{\per\second}$. We compare the models' predictions in \cref{fig:simpShear_shortExposure} over the exposure time $\tau = x / U$ of cells moving along the top wall.
Because of the short exposure time, the total deformation in \cref{fig:simpShear_shortExposure_l1} is lower than in the previous configuration. This is a feature of strain-based hemolysis models, which take into account the viscoelastic behavior of the cell membrane. 
Due to the higher levels of shear stress, the initial alignment of the cell with the tank-treading orientation in \cref{fig:simpShear_shortExposure_phi} happens faster than in the previous configuration. In particular, it still takes up only a miniscule fraction of the total exposure time. As a result, the \ac{TTM} is again able to predict orientation and deformation accurately in the relevant timescales, as evidenced by \cref{fig:simpShear_shortExposure_l1}.
In contrast, the alignment with the tank-treading state is much slower in the simplified mode. As a result, this model fails to predict any significant deformation over this exposure time.
}
\bigskip\par
Overall, this test case \replaced{verifies}{validates} the full-order Eulerian reformulation~\labelcref{eq:model_full-order}, which agrees perfectly with the Lagrangian model~\labelcref{eq:model_arora}. The \ac{TTM}~\labelcref{eq:model_tt} only deviates slightly for the initial orientation, but agrees for the deformation. Since deformation is the primary quantity of interest, the \ac{TTM} is a good approximation for the full-order model. 
The simplified model~\labelcref{eq:model_pauli}, on the other hand, deviates significantly for the transient cell deformation. This demonstrates the effects of the lower rotation rate. 
\subsection{Rotating shear}%
\label{sec:rotating_shear}%
In order to explore transient cell behavior in more detail, we consider two-dimensional circular Couette flow.
This test case is visualized in \cref{fig:rotShear_setup}. We choose $R = 0.701 \, \unit{\centi\metre}$, $\Delta R = 0.0001 \, \unit{\centi\metre}$ and counter-rotating walls with $\omega_\mathrm{i} = 1 \, \unit{\radian\per\second}$, $\omega_\mathrm{o} = -1 \, \unit{\radian\per\second}$. This system produces a nearly constant shear rate of $G_\mathrm{f} = 14{,}021 \, \unit{\per\second}$ across the domain. We define a non-dimensionalized radial coordinate $r^* = \frac{r - R}{\Delta R} \in [0,1]$. The domain is discretized with 8 elements in radial direction and 3560 elements in azimuthal direction. Due to the two-dimensional nature of the problem, the Taylor-Couette instability does not affect the flow. We simulate all Eulerian morphology models on this domain up to steady state. 
\par
The pathlines are closed circles with $r^* = \mathrm{const}$. Along each pathline, a cell experiences simple shear flow with a rotating direction of shear: 
\begin{equation*}
    \grad\vel(t) = 
    \rotMat{z}(\omega t) 
    \,
    \begin{bmatrix}
        0 & G_\mathrm{f} & 0\\
        0 & 0 & 0\\
        0 & 0 & 0
    \end{bmatrix} 
    \rotMat{z}\transp(\omega t) \, , \qquad
    \rotMat{z}(\varphi) =
    \begin{bmatrix}
        \cos\varphi & -\sin\varphi & 0 \\
        \sin\varphi & \cos\varphi & 0 \\
        0 & 0 & 1
    \end{bmatrix} \, .
\end{equation*}
The angular frequency depends on the radius of the pathline:
\begin{equation*}
    \omega \left( r^* \right) = 
    \omega_\mathrm{i} 
    + \left( \omega_\mathrm{o} - \omega_\mathrm{i} \right) 
    \left( 2 r^* - 1 \right) \, .
\end{equation*}
We evaluate the Arora model for three radii $r^* \in \{ 0, 0.5, 1 \}$ by integrating the Lagrangian morphology model along the circular pathlines up until steady state.
\par
\input{fig_rotatingShear.tex}In \cref{fig:rotShear_Geff}, we compare the different morphology models. With the Lagrangian Arora model, the effective shear rate $G_\mathrm{eff}$ is nearly constant across the range of $r^*$ and corresponds to fluid shear $G_\mathrm{f}$. As cells travel along their pathlines, they constantly realign themselves with respect to the rotating direction of shear. In the Arora model, this realignment occurs so rapidly that cells practically experience only simple shear. For simple shear, $G_\mathrm{eff} = G_\mathrm{f}$ holds in steady state, as discussed in \cref{sec:simple_shear}. 
\par
Out of the Eulerian models, the full-order model again agrees perfectly with the Lagrangian model. The \ac{TTM} agrees very well with the full-order model for these parameters.
The simplified model, on the other hand, deviates significantly from the full-order model. It underpredicts cell deformation at the inner wall and overpredicts it at the outer wall. It agrees with the full-order model only at the centerline $r^* = 0.5$. This is due to the lower rotation rate of cells in the simplified model, which causes a phase lag in their orientation as they travel along the circular pathline. At the inner wall, this phase lag causes the major axis to be less aligned with the principal axis of strain, leading to lower deformation. At the outer wall, the phase lag causes the major axis to be more aligned with the principal axis of strain, leading to higher deformation. At the centerline, cells are at rest, so they do not experience a rotating direction of shear, $\omega(r^* = 0.5) = 0$. This is equivalent to simple shear, where the simplified model agrees with the full-order model at steady state. 
\par
This test case serves as further \replaced{verification}{validation} for the Eulerian full-order model and the \ac{TTM}, both of which agree with the Lagrangian Arora model. Furthermore, it demonstrates that the simplified model deviates from the full-order model not only in transient behavior (see \cref{sec:simple_shear}), but also at steady state. In general, the simplified model is not suitable for any flows with non-constant gradients along its pathlines, as this requires cells to realign. This realignment incurs modeling errors due to the lower rotation rate in the simplified model. As this test case demonstrates, these errors can lead to both overprediction and underprediction of effective shear rate. 
\subsection{Square stirrer}
\label{sec:pauli_stir}
In order to test the morphology models' performance in more complex flows, we consider a two-dimensional square domain with a stirrer. The geometry and mesh are identical to those presented in previous works~\cite{pauli_stabilized_2016,hasler_variational_2019}.
We set a stirrer frequency of $\omega_\mathrm{stir} = 100 \, \unit{\radian\per\second}$ and use the \ac{MRF} approach for the rotating stirrer. The interface is visualized as a dashed line in \cref{fig:pauliStir_simpl}. The computational domain is discretized with $92{,}262$ triangular elements. The initial condition is an undeformed cell. We simulate the Eulerian morphology models up to steady state. 
\deleted{
The models are compared with respect to computational performance in Tab. 1. The full-order model is the slowest, as its time step size is severely restricted by the small timescale of the cell rotation (cf. Section 4.1). This limits its applicability to simple two-dimensional flows, and it becomes prohibitively expensive for three-dimensional flows. It is thus not suited for realistic blood pump geometries. The simplified model is not affected by this issue to the same degree, as cells rotate at a slower rate. The time step limitation is thus not as severe. The \ac{TTM} is more robust, since it does not resolve cell rotation explicitly. This allows for larger time steps and thus significantly faster computation. 
}
\input{fig_pauliStir.tex}
\par
The model predictions are compared in \cref{fig:pauliStir}. The solution in \cref{fig:pauliStir_simpl} matches previous results using the simplified model~\cite{hasler_variational_2019}. In comparison with the full-order model in \cref{fig:pauliStir_full}, however, it deviates significantly. In particular, it underpredicts effective shear rate close to the stirrer by two orders of magnitude. The \ac{TTM} in \cref{fig:pauliStir_tt} agrees well with the full-order model, but lacks some of the finer details of the solution in the region close to the stirrer. Finally, the three models are compared along the profile line $\xi$ in \cref{fig:pauliStir_profLine}. The results for the simplified model agree qualitatively with those presented by Ha\ss ler et al.~\cite{hasler_variational_2019}. However, this agrees with the full-order model only close to the outer walls. In that region, flow velocities are lower, so cells do not have to respond to changing velocity gradients as quickly. The simplified model is thus able to predict the effective shear rate accurately. Closer to the center, cells rotate more quickly, so the modeling error due to the lower rotation rate of the simplified model becomes more significant. This causes the underprediction in that region. Conversely, the \ac{TTM} agrees well with the full-order model across the whole domain.

\subsection{Simple blood pump}
\label{sec:simple_pump}

The final test case involves a three-dimensional geometry. \added{It was designed by Pauli et al.~\cite{pauli_transient_2013} to represent a simplified version of the FDA benchmark pump~\cite{ponnaluri_results_2023}. In particular, it is lacking the central cone element of the impeller.}  The domain is discretized with a total of $6{,}618{,}708$ tetrahedral elements.
We simulate fluid flow through the pump with the \ac{MRF} approach at a flow rate of $0.5 \, \unit{\litre\per\minute}$, with a rotation rate of $1000 \, \mathrm{rpm}$. 
\added{The Reynolds numbers in the inflow tube and impeller region are $\mathrm{Re}_\mathrm{in} = \rho \bar{u}_\mathrm{in} d_\mathrm{in} / \mu = 320$ and $\mathrm{Re}_\mathrm{imp} = \rho \omega_{\mathrm{rot}} r_\mathrm{imp}^2 / \mu = 25{,}712$, respectively. Since this corresponds to a laminar flow regime \cite{fraserUseComputationalFluid2011}, we do not employ a turbulence model.}
For the Eulerian morphology models, we set undeformed cells as initial condition and as boundary condition at the inlet. We employ the \acf{TTLM} for additional robustness. 
\added{We perform unsteady simulations until we reach a quasi-steady state, i.e., the flow field and cell deformation do not change significantly over time.}
\begin{table}
    \centering
    \begin{tabular}{lr}
        \toprule 
        Simulation & Time \\
        \midrule
        \ac{CFD} (MRF) & $20 \, \mathrm{h}$ \\
        \added{Full-order morphology (est.)} & \added{$4000 \, \mathrm{h}$} \\
        Simplified morphology & $45 \, \mathrm{min}$ \\
        \ac{TTLM} & \replaced{$4 \, \mathrm{min}$}{$8 \, \mathrm{min}$} \\
        \bottomrule
    \end{tabular}
    \caption{Execution times for the simple pump test case on 192 cores.}
    \label{tab:simplePump_time}
\end{table}
\par
\added{The execution times to obtain the respective solutions are listed in \cref{tab:simplePump_time}. 
The CFD simulation requires 833 time steps of size $\Delta t = 5 \cdot 10^{-4}$ to produce a steady solution. Each time step takes $86 \, \mathrm{s}$ on 192 cores. In this complex flow, cells exhibit unsteady tumbling behavior even in quasi-steady state. The two morphology models \labelcref{eq:model_full-order,eq:model_pauli} require us to resolve this rotation.
For the simplified morphology model, we require 60 time steps, which take 45 seconds each on 192 cores. 
As the rotation rate of cells in the full-order model is larger by a factor of $1 / f_3$ (comp. \cref{eq:om_explicit,eq:om_pauli}), we would require at least 142{,}000 time steps for the full-order model. Each step takes more than 100 seconds due to the eigenvector operations, leading us to an estimated total execution time of at least 4000 hours, two orders of magnitude larger than the CFD simulation.
For this reason, we do not perform the Eulerian full-order morphology simulation for this test case.
Finally, the \ac{TTLM} does not need to resolve cell rotation and thus only requires 30 time steps. Additionally, each time step only takes $7 \, \unit{\second}$ due to the lower number of degrees of freedom.}
\input{fig_simplePump-planeComp.tex}
\par
\Cref{fig:simplePump_planeComp} shows the effective shear rate on a plane parallel to the impeller disk. The plane is located $0.5 \, \unit{\milli\metre}$ above the blades. The simplified model in \cref{fig:simplePump_planeComp_pauli} again underpredicts the effective shear rate severely compared to the \ac{TTLM} in \cref{fig:simplePump_planeComp_tank}. The relative error $\delta = | G_\mathrm{eff}^\mathrm{simpl} - G_\mathrm{eff}^\mathrm{TTLM} | /  G_\mathrm{eff}^\mathrm{TTLM}$ is shown in \cref{fig:simplePump_planeComp_delta}. Close to the outer wall and close to the inflow in the center, the two models agree. This can be explained by the same reasoning as in the square stirrer test case (cf. \cref{sec:pauli_stir}): The simplified model is able to predict the effective shear rate accurately in regions with low flow velocity, where cells do not have to respond to changing velocity gradients as quickly. Everywhere else, the simplified model underpredicts the effective shear rate by up to 98\%.
\par
\input{fig_simplePump-pathlinesCoverage.tex}
\input{fig_simplePump-pathlines.tex}
For \replaced{verification}{validation}, we compare the Eulerian and Lagrangian model formulations as outlined in \cref{sec:lagrangian_verification}.
Four particular pathlines are selected and visualized in \cref{fig:pathlines_coverage}. The pathlines provide good coverage of the area around the impeller. However, only one of the four pathlines exits the domain through the outlet. The other pathlines terminate at walls or at the impeller. This is an artifact of the discrete nature of the velocity field and the explicit time stepping scheme used to integrate the pathlines. It presents a principal limitation of the Lagrangian approach, as it is not possible to track every cell through the whole domain, especially in complex geometries, where cells can get trapped in recirculation zones. This limitation does not apply to the Eulerian models, as they are solved on the whole domain.
\par
The model results along these particular pathlines are given in \cref{fig:pathlines}. The two Lagrangian solutions for Arora model and \ac{TTM} are practically identical, underlining the validity of the \ac{TTM} to approximate the full-order behavior. Out of the Eulerian models, the simplified model again underpredicts the effective shear rate significantly. For systems with higher rotation rates of the impeller, we expect this discrepancy to grow further, as cell rotation becomes more significant.
The Eulerian formulation of the \ac{TTM} agrees overall well with its Lagrangian formulation. The Eulerian formulation is more diffusive due to the finite element discretization in space. The Lagrangian formulation does not require any mesh, as its discretization is only governed by the integration step size along its pathline. This causes some local discrepancies between the solutions. However, if we integrate empirical hemolysis models along a larger set of 36 pathlines, the averaged results agree within $5 \, \%$. Overall, this comparison demonstrates that the Eulerian formulation of the \ac{TTM} is a valid approximation of the Lagrangian Arora model for the given application.

%% file: fig_simpleShear.tex
\begin{figure}
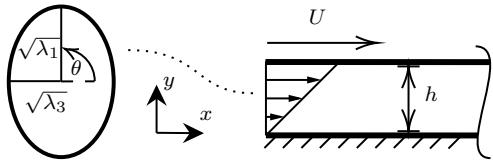
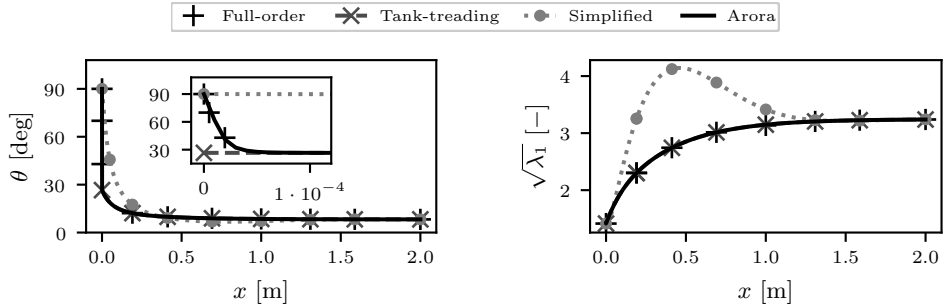

    \begin{subfigure}{\textwidth}
        \centering
        \importpgf{results_simpleShear}{setup}
        \caption{Test case configuration}
        \label{fig:simpShear_setup}
    \end{subfigure}
    \\[1em]
    \begin{subfigure}{\textwidth}
        \centering
        \importpgf{results_simpleShear}{legend}
    \end{subfigure}
    \centering
    \begin{subfigure}{0.48\textwidth}
        \importpgf{results_simpleShear}{phi}
        \caption{Angle of major axis to flow direction.}
        \label{fig:simpShear_phi}
    \end{subfigure}
    \begin{subfigure}{0.48\textwidth}
        \importpgf{results_simpleShear}{l1}
        \caption{Major axis length.}
        \label{fig:simpShear_l1}
    \end{subfigure}
    \caption{Comparison of morphology models under simple shear.}
    \label{fig:simpShear}
\end{figure}

%% file: fig_simpleShear_shortExposure.tex
\begin{figure}
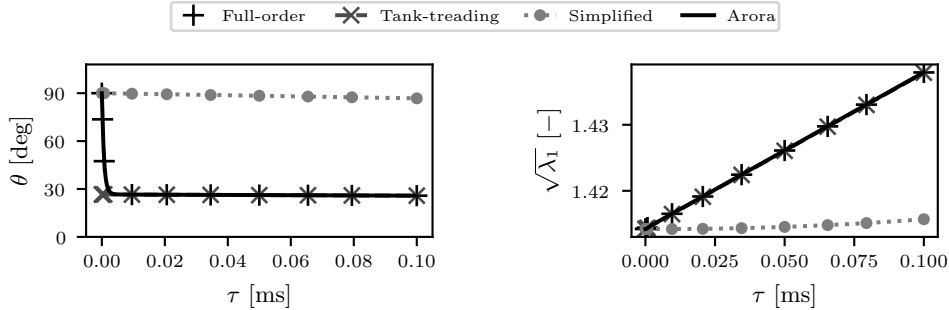

    \begin{subfigure}{\textwidth}
        \centering
        \importpgf{results_simpleShear}{l1_shortExposure_legend}
    \end{subfigure}
    \centering
    \begin{subfigure}{0.48\textwidth}
        \importpgf{results_simpleShear}{phi_shortExposure}
        \caption{Angle of major axis to flow direction.}
        \label{fig:simpShear_shortExposure_phi}
    \end{subfigure}
    \begin{subfigure}{0.48\textwidth}
        \importpgf{results_simpleShear}{l1_shortExposure}
        \caption{Major axis length.}
        \label{fig:simpShear_shortExposure_l1}
    \end{subfigure}
    \caption{Comparison of morphology models under short exposure times $\tau = x / U$.}
    \label{fig:simpShear_shortExposure}
\end{figure}

%% file: fig_rotatingShear.tex
\begin{figure}
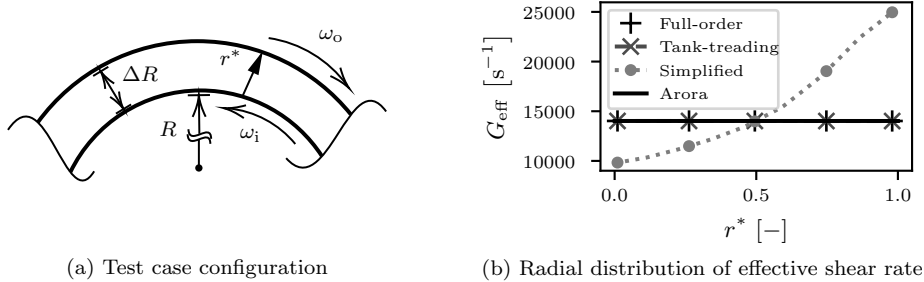

    \centering
    \begin{subfigure}[t]{0.48\textwidth}
        \centering
        \vskip 1.6em
        \importpgf{results_rotatingShear}{setup}
        \caption{Test case configuration}
        \label{fig:rotShear_setup}
    \end{subfigure}
    \begin{subfigure}[t]{0.48\textwidth}
        \centering
        \vskip 0pt
        \importpgf{results_rotatingShear}{Geff_counterRot}
        \vskip -1em
        \caption{Radial distribution of effective shear rate}
        \label{fig:rotShear_Geff}
    \end{subfigure}
    \caption{Comparison of morphology models under rotating shear.}
    \label{fig:rotShear}
\end{figure}

%% file: fig_pauliStir.tex
\begin{figure}
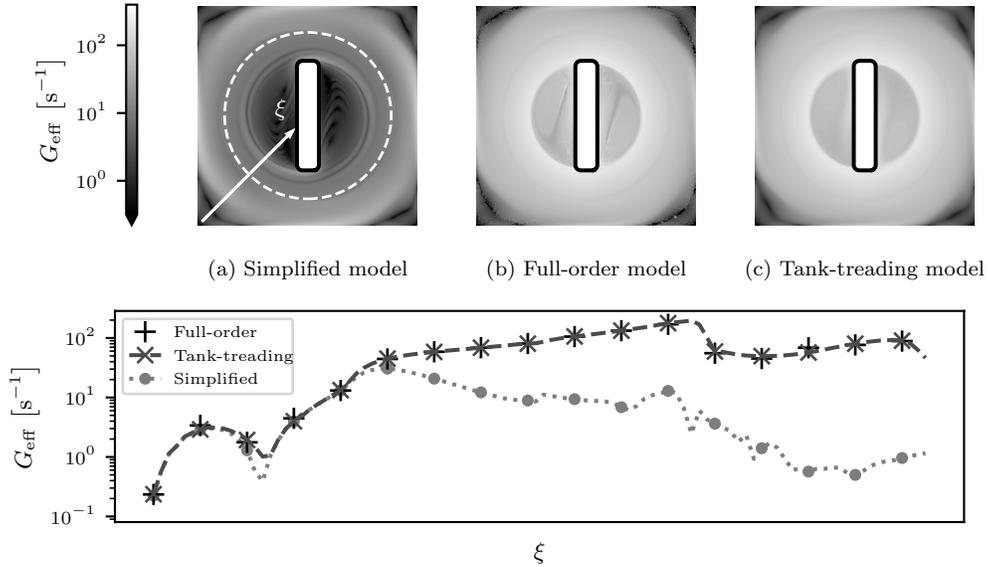

    \centering
    \begin{subfigure}[t]{0.16\textwidth}
        \importpgf{results_pauliStir}{cb}
    \end{subfigure}
    \begin{subfigure}[t]{0.26\textwidth}
        \centering
        \importpgf{results_pauliStir}{Geff_pauli}
        \vspace{-3ex}
        \caption{Simplified model}
        \label{fig:pauliStir_simpl}
    \end{subfigure}
    \begin{subfigure}[t]{0.26\textwidth}
        \centering
        \importpgf{results_pauliStir}{Geff_full}
        \vspace{-3ex}
        \caption{Full-order model}
        \label{fig:pauliStir_full}
    \end{subfigure}
    \begin{subfigure}[t]{0.26\textwidth}
        \centering
        \importpgf{results_pauliStir}{Geff_tt}
        \vspace{-3ex}
        \caption{Tank-treading model}
        \label{fig:pauliStir_tt}
    \end{subfigure}
    \\
    \begin{subfigure}[t]{\textwidth}
        \centering
        \importpgf{results_pauliStir}{G_profLine_1}
        \vspace{-2ex}
        \caption{Comparison between morphology models along line from lower left corner to center (see \cref{fig:pauliStir_simpl}).}
        \label{fig:pauliStir_profLine}
    \end{subfigure}
    \caption{Morphology results for square stirrer}
    \label{fig:pauliStir}
\end{figure}

%% file: fig_simplePump-planeComp.tex
\begin{figure}[ht]
    \begin{subfigure}{0.48\textwidth}
        \includegraphics[width=0.98\textwidth]{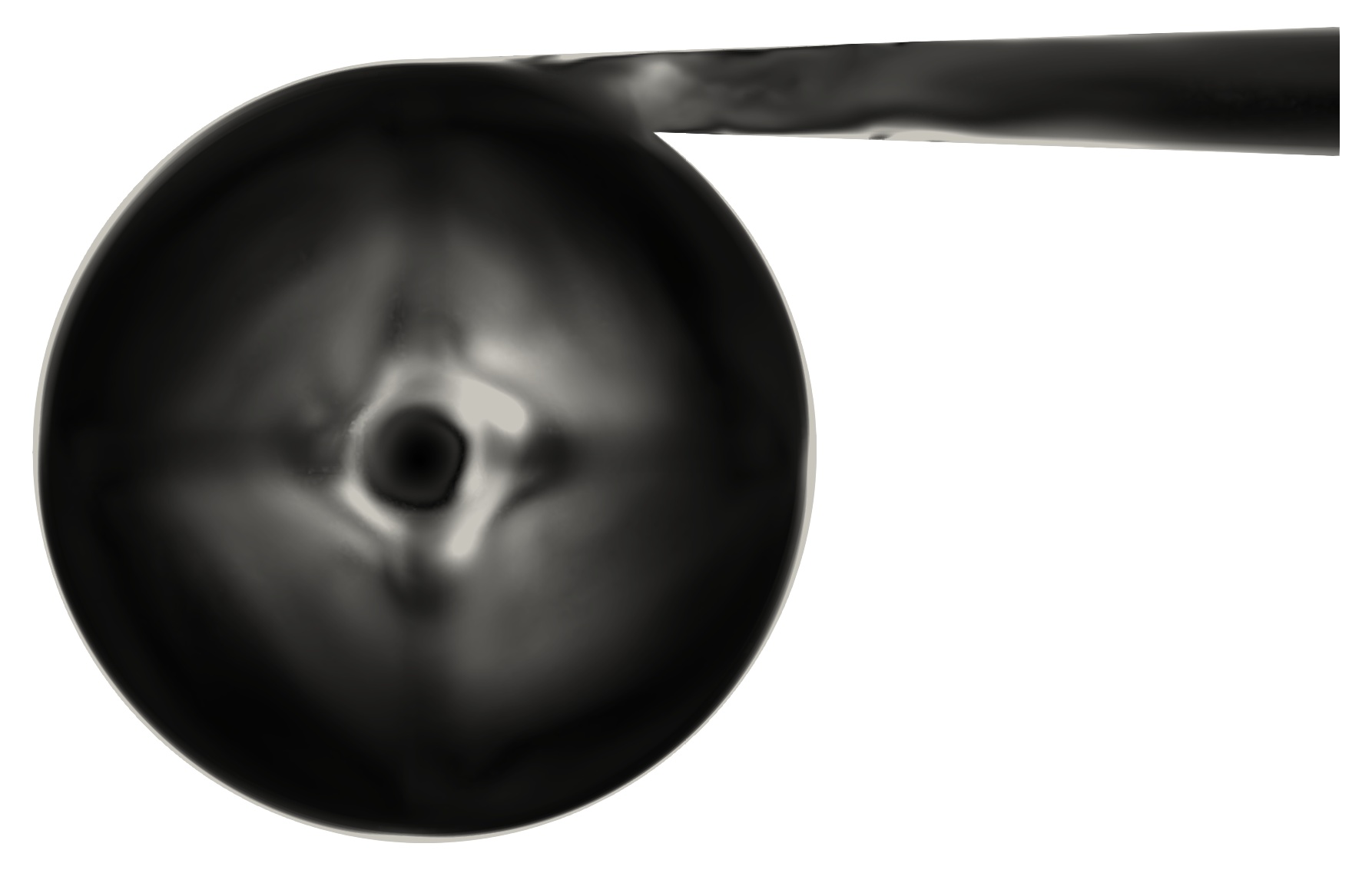}
        \hspace{-7em}
        \importpgf{results_simplePump}{cb_250}
        \caption{Simplified model}
        \label{fig:simplePump_planeComp_pauli}
    \end{subfigure}
    \begin{subfigure}{0.48\textwidth}
        \includegraphics[width=0.98\textwidth]{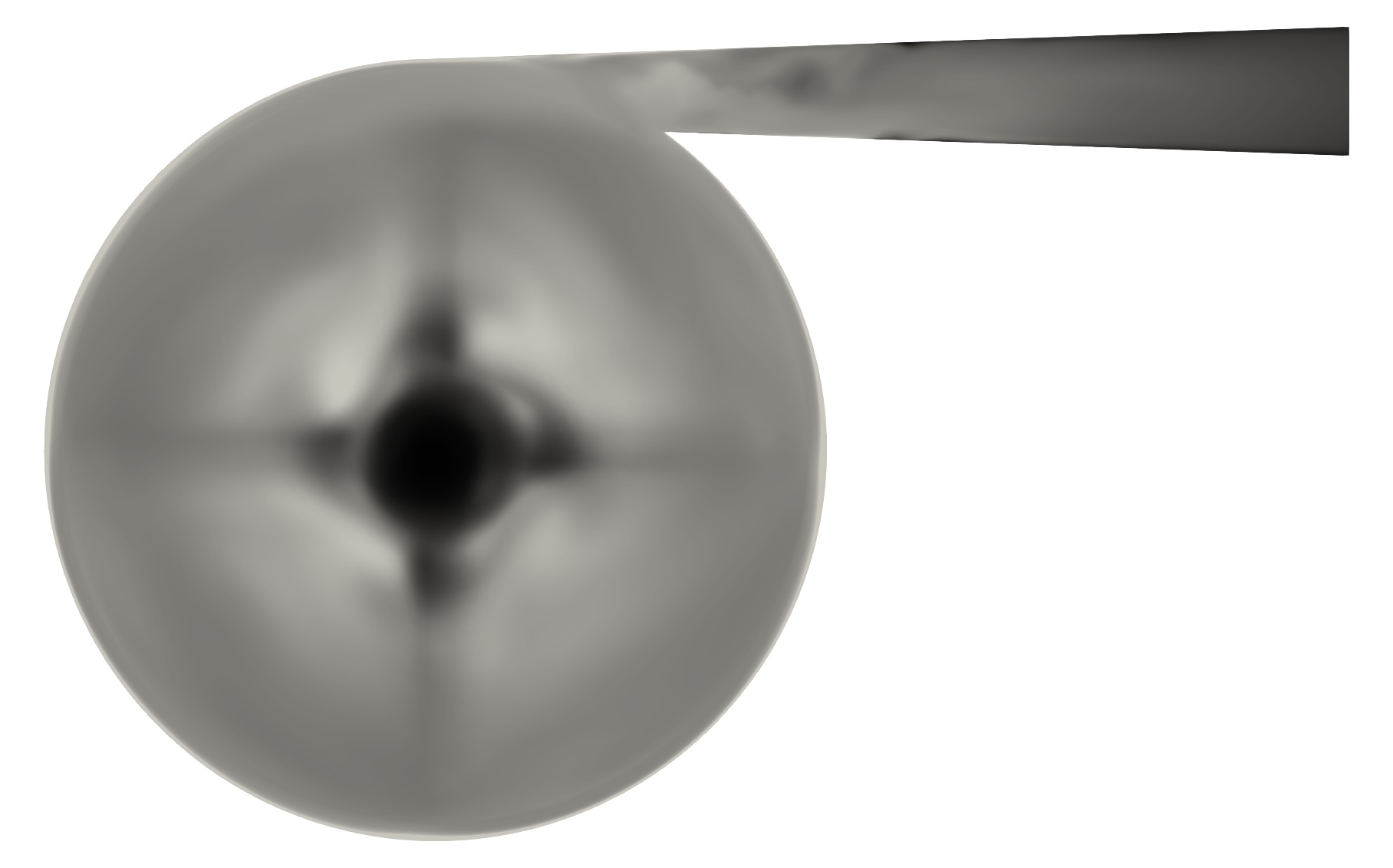}
        \hspace{-7em}
        \importpgf{results_simplePump}{cb_1000}
        \caption{Tank-treading model}
        \label{fig:simplePump_planeComp_tank}
    \end{subfigure}
    \centering
    \begin{subfigure}{0.65\textwidth}
        \includegraphics[width=0.98\textwidth]{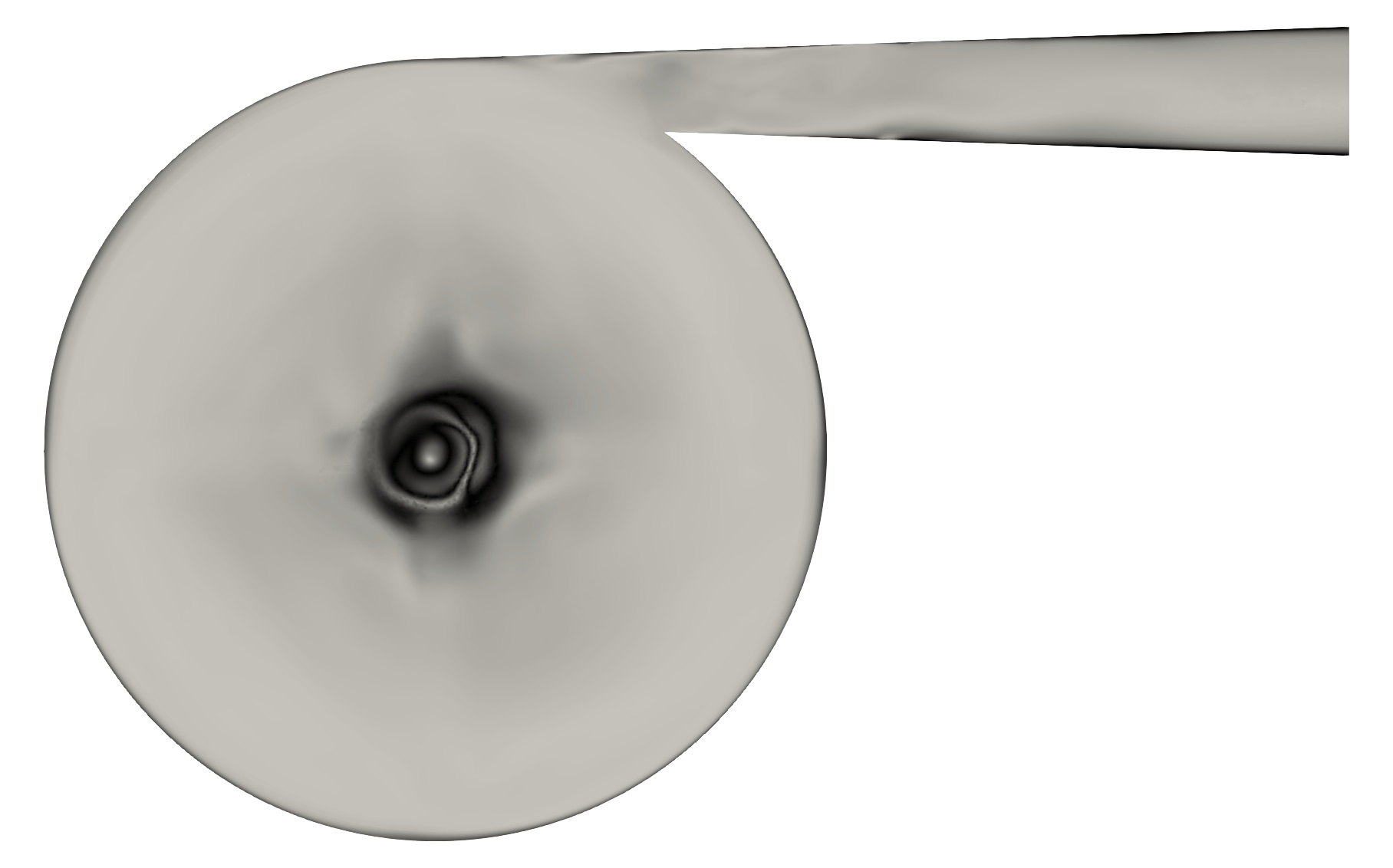}
        \hspace{-7em}
        \importpgf{results_simplePump}{cb_delta}
        \caption{Relative error of simplified model}
        \label{fig:simplePump_planeComp_delta}
    \end{subfigure}
    \caption{Effective shear rate in simplified three-dimensional blood pump in the plane $5 \, \unit{\milli\meter}$ above the impeller blades.}
    \label{fig:simplePump_planeComp}
\end{figure}

%% file: fig_simplePump-pathlinesCoverage.tex
\begin{figure}[p]
    \centering
    \includegraphics[width=0.8\textwidth]{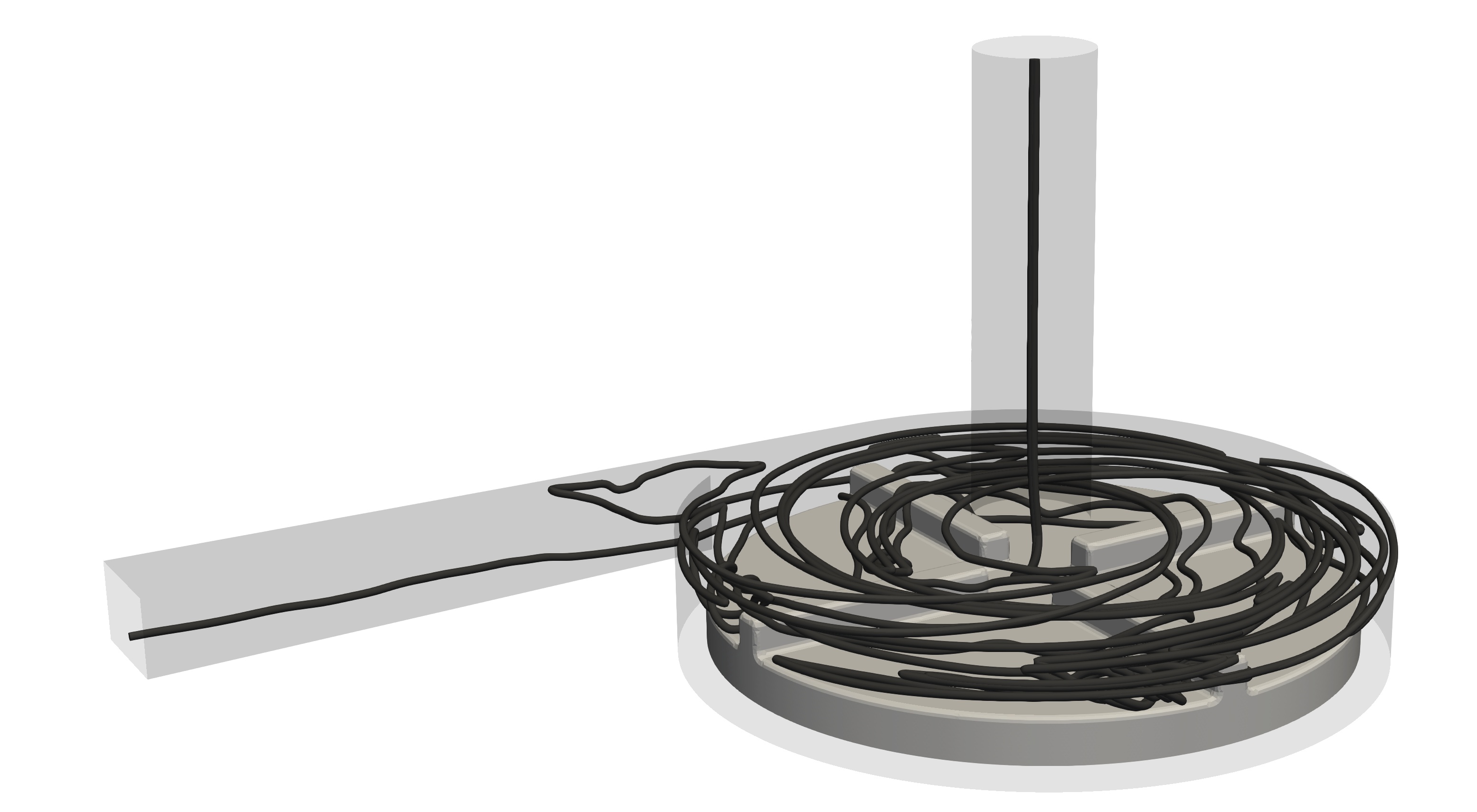}
    \caption{Selected pathlines in the simple blood pump.}
    \label{fig:pathlines_coverage}
\end{figure}

%% file: fig_simplePump-pathlines.tex
\begin{figure}[p]
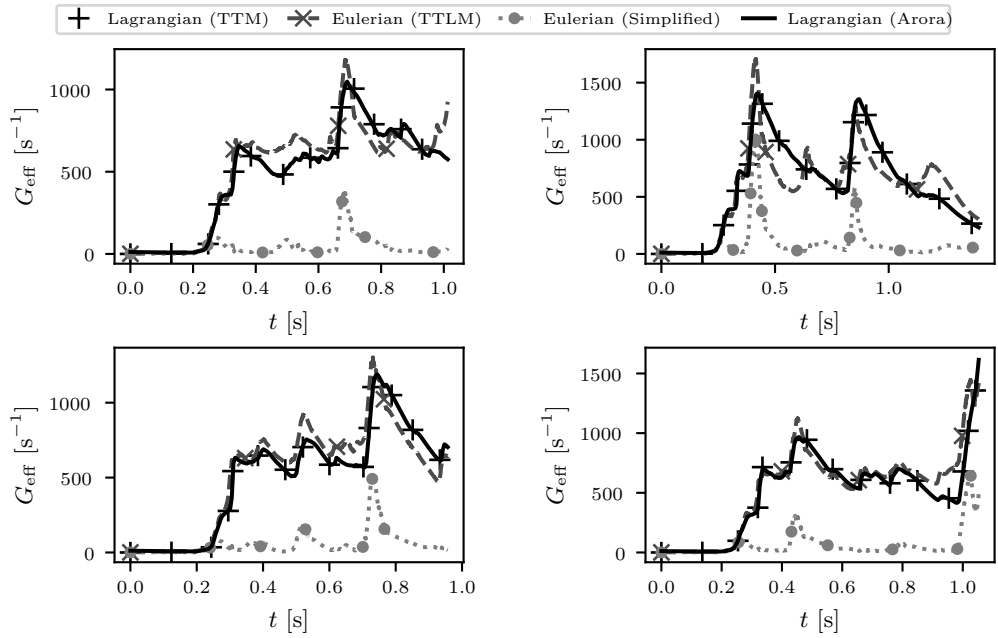

    \begin{subfigure}{\textwidth}
        \centering
        \importpgf{results_simplePump}{legend}
    \end{subfigure}
    \begin{subfigure}{0.485\textwidth}
        \importpgf{results_simplePump}{pl-0}
        \label{fig:pathlines_0}
    \end{subfigure}
    \hfill
    \begin{subfigure}{0.485\textwidth}
        \importpgf{results_simplePump}{pl-1}
        \label{fig:pathlines_1}
    \end{subfigure}
    \begin{subfigure}{0.485\textwidth}
        \importpgf{results_simplePump}{pl-3}
        \label{fig:pathlines_2}
    \end{subfigure}
    \hfill
    \begin{subfigure}{0.485\textwidth}
        \importpgf{results_simplePump}{pl-4}
        \label{fig:pathlines_3}
    \end{subfigure}
    \caption{Eulerian and Lagrangian model formulations compared along four sample pathlines in the simple blood pump.}
    \label{fig:pathlines}
\end{figure}

%% file: sections_conclusion.tex
\section{Conclusion and Outlook}
\label{sec:conclusion}

\deleted{
Hemolysis modeling is an integral part of the design process of blood-contacting medical devices. But as of now, no standard way of modeling this phenomenon has been established. Instead, the landscape of models stretches from relatively simple Eulerian stress-based models to complex and less practical Lagrangian strain-based models. With this work, we aim to provide some middle ground by combining the unique advantages of both approaches into the Eulerian \acf{TTM}. 
Compared to Eulerian stress-based models, the \ac{TTM} is able to capture the viscoelastic behavior of the cell membrane by resolving cell deformation time. Compared to Lagrangian strain-based models, the \ac{TTM} is computationally more efficient and does not require tracking individual cells. Compared to a previous Eulerian strain-based model~\cite{pauli_transient_2013}, the \ac{TTM} more accurately captures the original Lagrangian model behavior. This has been shown on the basis of a selection of benchmark flows, including simple shear, rotating shear, and two more complex rotational flows representative of \acp{VAD}. 
}%
\added{
We have demonstrated the capabilities of our new Eulerian strain-based hemolysis model, the \acf{TTM}, in a variety of benchmark flows.
Presently, its largest limitation is the underlying assumption of ellipsoidal cell shape. 
More complex dynamics and morphologies have been shown to occur in various flow conditions~\cite{fedosov_deformation_2014,lanotte_red_2016,mauer_flow-induced_2018,ezzeldin_strain-based_2015}.}
We argue that the present model nevertheless provides practical predictions for two reasons. 
First, fully resolving small-scale cell behavior requires a more complex structural model for the cell membrane. Such models currently only allow for simulation of small ensembles of cells, many orders of magnitude away from the total number of cells in realistic configurations. They are thus not suitable for the design process of \acp{VAD}.
Second, scale-resolving simulations by Lanotte et al.~\cite{lanotte_red_2016} at physiological \emph{hematocrit} (volume fraction of \acp{RBC}) and physiological plasma viscosities indicate that at high shear rates, elongated flattened cells with tank-treading dynamics dominate. Even though their shapes may feature irregularities compared with pure ellipsoids, orientation and characteristic deformation time should behave similarly. We thus expect the \ac{TTM} to still describe the deformation of such cells adequately.
\par
In the future, we will examine if the effects of the aforementioned morphology irregularities can be incorporated into the model phenomenologically, e.g., by adapting the model parameters $f_1, f_2, f_3$. 
Additionally, we are aiming to obtain experimental data for cell deformation. This will allow for model validation and provide a basis for the calibration of the model parameters, especially in light of new results for relaxation time~\cite{guglietta_loading_2021} and methods of varying the parameters~\cite{taglienti_reduced_2023}.
Finally, we intend to apply the \ac{TTM} to more realistic \ac{VAD} geometries and compare its hemolysis predictions with computational and empirical reference results, e.g., for the FDA benchmark blood pump~\cite{ponnaluri_results_2023}. 
In this context, we are also planning to investigate the effects of turbulence modeling and variable hematocrit on strain-based hemolysis predictions.
\par
Overall, we are confident that the \ac{TTM} represents an excellent compromise between accuracy and practicality. 
\added{%
In contrast to Eulerian stress-based models, the \ac{TTM} is able to capture the viscoelastic behavior of the cell membrane by resolving cell deformation time. Compared to Lagrangian strain-based models, the \ac{TTM} is computationally more efficient and does not require tracking individual cells. Compared to a previous Eulerian strain-based model~\cite{pauli_transient_2013}, the \ac{TTM} more accurately captures the original Lagrangian model behavior.
}%
\deleted{It provides a solid basis for understanding the effects of bulk flow on \acp{RBC}.}
These qualities will make it a valuable tool for the design process of future generations of medical devices.

%% file: sections_acknowledgements.tex
\section*{Acknowledgements}

This work was funded by the Deutsche Forschungsgemeinschaft (DFG, German Research Foundation) through grant 333849990/GRK2379 (IRTG Modern Inverse Problems). The authors gratefully acknowledge the computing time granted by the JARA Vergabegremium and provided on the JARA Partition part of the supercomputer CLAIX at RWTH Aachen University and the computing time provided to them on the high-performance computer Lichtenberg at the NHR Centers NHR4CES at TU Darmstadt. This is funded by the Federal Ministry of Education and Research, and the state governments participating on the basis of the resolutions of the GWK for national high performance computing at universities (www.nhr-verein.de/unsere-partner).

%% file: sections_rotation.tex
\section{On the definition of the rotation term}
\label{sec:rotation}

The left-hand side of the Arora model~\labelcref{eq:model_arora} is meant to represent a Jaumann derivative. The defining feature of a Jaumann derivative is that it is objective under rotation, i.e., in a purely rotational flow and neglecting shape recovery, it has to hold that
\begin{equation*}
    \oddt{\morph} - \comm{\Om}{\morph} = \zeroMat \, .
\end{equation*}
Again employing the spectral decomposition~\labelcref{eq:specDecomp} and noting that the eigenvalues are constant in the case of pure rotation, we find that
\begin{align*}
    \oddt{\morph} &= \oddt{\Qm} \Lamb \Qm\transp + \Qm \Lamb \oddt{\Qm\transp} \, , \\
    \comm{\Om}{\morph} &= \Om \Qm \Lamb \Qm\transp - \Qm \Lamb \Qm\transp \Om \, .
\end{align*}
Thus, objectivity is ensured if
\begin{equation*}
    \oddt{\Qm} = \Om \Qm 
    \qquad \text{and} \qquad
    \oddt{\Qm\transp} = \Qm\transp \Om \, .
\end{equation*}
These equations are only satisfied by the definition~\labelcref{eq:om}.
In contrast, Arora~\cite{d_arora_tensor-based_2004} employed the definition 
\begin{equation*}
    \Omt = - \oddt{\Qm\transp}\Qm = \Qm\transp \oddt{\Qm} = \Qm\transp \oddt{\Qm} \Qm \Qm\transp = \Qm\transp \Om \Qm \, ,
\end{equation*}
which corresponds to a change of basis to the coordinate system of the cell. In tensor notation, the two definitions are equivalent.

%% file: sections_implementation.tex
\section{Finite element code}
\label{sec:implementation}

\deleted{Having explained the theoretical foundations of our approach to Eulerian hemolysis simulation based on the flow and morphology problems, we now turn to its practical implementation by integrating it into our in-house multiphysics finite element code \texttt{XNS}.}
Our in-house multiphysics finite element code \texttt{XNS} is written in \texttt{Fortran} and parallelized by the in-house \texttt{EWD} communication library using the \ac{MPI}.
This allows us to exploit the potential of modern \ac{HPC} architectures, especially for large-scale problems. 
\texttt{XNS} supports semi-discrete as well as space-time finite element formulations~\cite{pauli_stabilized_2017} and entails a number of advanced methods for moving and deforming grids~\cite{hilgerNovelApproachFluidStructure2021,helmigCombiningBoundaryConformingFinite2021a,gonzalezSurfaceReconstructionVirtualRegionMesh2023a,keyVirtualRingShearSlip2018a}. It also provides capabilities to perform parametric model order reduction~\cite{keyModelOrderReduction2023,keyReducedFlowModel2021a} using the \texttt{RBniCS}\footnote{RBniCS — reduced order modeling in FEniCS, \url{https://www.rbnicsproject.org/}} library~\cite{hesthavenCertifiedReducedBasis2015}.
\bigskip\par
The design of \texttt{XNS} adheres to the \ac{OOP} paradigm. 
Here, we employ encapsulation to structure the code in a modular fashion, which comes with increased data security and promotes a more intuitive understanding of the code's design. In particular, the code is organized to separate the numerical discretization aspects, including computational mesh and time discretization, from the formulation of physical field equations and the incorporation of features such as stabilization schemes or discontinuity capturing. 
Furthermore, the design also facilitates coupling, e.g., between multiple field problems, which requires exchanging the corresponding solution fields. Here, \texttt{XNS} provides functionality to couple field problems both weakly and strongly, according to one-way or mutual dependencies.
\par
The \ac{OOP} approach also enables the usage of inheritance and polymorphism. 
On the one hand, inheritance is leveraged throughout the code to create a hierarchy of classes, allowing the derived classes to inherit properties and behaviors from parent classes. Thereby, common functionality can be efficiently implemented by reusing and sharing existing code. 
For example, the distribution of temperature or concentration can be governed by the same scalar transport equation, but with different material parameters or source terms. Using the concept of inheritance, this unified equation structure allows us to encapsulate the common functionality while still addressing the specific variations.  
\par
On the other hand, polymorphism is implemented, enabling objects of different classes to be treated as objects of a common superclass.
As an illustrative example, the use of polymorphism allows the assembly process in the finite element procedure to be implemented only once for different field problems derived from a common superclass. This design approach allows seamless interchangeability and coupling of different problem types and avoids code duplication.
\bigskip\par
The above concepts are also particularly useful in the context of this work.
Due to the modular design, it was straightforward to create an additional module for the morphology problem and integrate it into the existing framework of field problems. 
\par
All field problems are derived from a common parent class \texttt{FieldProblemT}. The flow problem is implemented as a subclass \texttt{INSProblemT}.
For the different Eulerian models from \cref{sec:model}, we followed the concept of inheritance to reflect their common properties. We introduced a new parent class \texttt{MorphologyProblemT} related to the generic formulation~\labelcref{eq:morphology_weak_form}. From this, we derived classes for the specific models that differ in number of degrees of freeedom and in definition of the source term. 
\texttt{MorphologyTensorProblemT} is the parent class for the morphology models that involve the morphology tensor $\morph$. The full-order model~\labelcref{eq:model_full-order} and the simplified model~\labelcref{eq:model_pauli} are implemented as subclasses \texttt{MorphologyFullOrderT} and \texttt{MorphologySimplifiedT}, respectively.
\texttt{MorphologyEigProblemT} implements the problems that operate directly on the eigenvalues $\lambda_i$ of $\morph$. The \ac{TTM} from \cref{eq:model_tt} and the \ac{TTLM} from \cref{eq:model_ttl} are implemented as subclasses \texttt{TankTreadingT} and \texttt{TankTreadingLogarithmicT}, respectively.
\Cref{fig:implementationMorphology} gives an overview of the classes and their inheritance.
\input{fig_inheritance.tex}
\par
As has been mentioned above, the flow and morphology problems are coupled in a weak manner. Thereby, the solution of the flow problem, i.e., the velocity field and its gradient, is used in the morphology problem.

%% file: fig_inheritance.tex
\begin{figure}
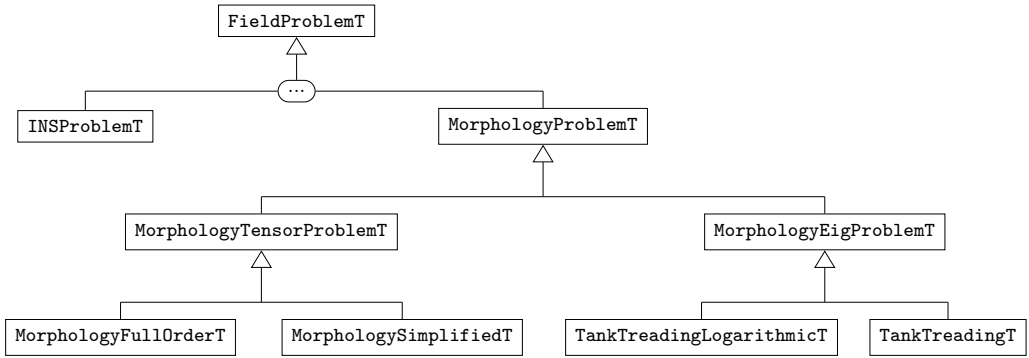

    \centering
    \begin{adjustbox}{width=\textwidth}
        \importtikz{implementation}{classes_vertical}
    \end{adjustbox}
    \caption{Inheritance structure of Eulerian field problems in \texttt{XNS}.}
    \label{fig:implementationMorphology}
\end{figure}